\documentclass[twocolumn, prx, superscriptaddress,notitlepage]{revtex4-2}
\usepackage{graphicx}
\usepackage{dcolumn}
\usepackage{bm}
\usepackage[usenames,dvipsnames]{color}
\usepackage[most]{tcolorbox}
\usepackage{multirow}
\usepackage{gensymb}
\usepackage[normalem]{ulem}
\usepackage{CJK}
\usepackage{comment}
\usepackage[colorlinks, linkcolor=blue,anchorcolor=blue,citecolor=blue,urlcolor=blue]{hyperref}
\usepackage{amssymb}
\usepackage{pifont}
\usepackage{physics}
\usepackage{natbib}
\usepackage{xcolor}

\usepackage{color,soul}

\begin{document}
\begin{CJK*}{UTF8}{}
\title{
Characterizing temperature and strain variations with  qubit ensembles for their robust coherence protection
}

\author{Guoqing Wang \CJKfamily{gbsn}(王国庆)}
\thanks{These authors contributed equally.}
\affiliation{
   Research Laboratory of Electronics, Massachusetts Institute of Technology, Cambridge, MA 02139, USA}
\affiliation{
   Department of Nuclear Science and Engineering, Massachusetts Institute of Technology, Cambridge, MA 02139, USA}

\author{Ariel Rebekah Barr}
\thanks{These authors contributed equally.}
\affiliation{Department of Materials Science and Engineering, Massachusetts Institute of Technology, MA 02139, USA}

\author{Hao Tang}
\thanks{These authors contributed equally.}
\affiliation{Department of Materials Science and Engineering, Massachusetts Institute of Technology, MA 02139, USA}

\author{Mo Chen} 
\affiliation{
   Research Laboratory of Electronics, Massachusetts Institute of Technology, Cambridge, MA 02139, USA}
\affiliation{Thomas J. Watson, Sr., Laboratory of Applied Physics, California Institute of Technology, Pasadena, CA 91125, USA}
\affiliation{Institute for Quantum Information and Matter, California Institute of Technology, Pasadena, CA 91125, USA}
   
\author{Changhao Li} 
\affiliation{
   Research Laboratory of Electronics, Massachusetts Institute of Technology, Cambridge, MA 02139, USA}
\affiliation{
   Department of Nuclear Science and Engineering, Massachusetts Institute of Technology, Cambridge, MA 02139, USA}

\author{Haowei Xu}
\affiliation{
   Department of Nuclear Science and Engineering, Massachusetts Institute of Technology, Cambridge, MA 02139, USA}

\author{Ju Li}\email[]{liju@mit.edu}
\affiliation{
   Department of Nuclear Science and Engineering, Massachusetts Institute of Technology, Cambridge, MA 02139, USA}
\affiliation{Department of Materials Science and Engineering, Massachusetts Institute of Technology, MA 02139, USA}

\author{Paola Cappellaro}\email[]{pcappell@mit.edu}
\affiliation{
   Research Laboratory of Electronics, Massachusetts Institute of Technology, Cambridge, MA 02139, USA}
\affiliation{
   Department of Nuclear Science and Engineering, Massachusetts Institute of Technology, Cambridge, MA 02139, USA}
\affiliation{Department of Physics, Massachusetts Institute of Technology, Cambridge, MA 02139, USA}

\begin{abstract}
Solid-state spin defects, especially nuclear spins with potentially achievable long coherence times, are compelling candidates for quantum memories and sensors. However, their current performances are still limited by the decoherence due to the variation of their intrinsic quadrupole and hyperfine interactions.
We propose an \textit{unbalanced echo} to overcome this challenge by using a second spin to refocus the variation of these interactions, which preserves the quantum information stored in the free evolution. 
The unbalanced echo can be used to probe the temperature and strain distribution in materials.
Experimentally, we demonstrate a 20-fold $T_2^*$ coherence time increase in an ensemble of $\sim10^{10}$ nuclear spins in diamond.  Theoretically, we develop first-principles methods to predict these interaction variations and reveal their correlation in large temperature and strain ranges. We numerically show that our method can refocus stronger noise variations than our current experiments and achieves a 400-fold coherence improvement for a 25~K temperature inhomogeneity.
\end{abstract}

\maketitle

\end{CJK*}

Solid-state spin defects are promising quantum information processing platforms for sensing, communication and computing due to their favorable optical, spin, and charge properties~\cite{wolfowicz_quantum_2021}. For example,  color centers in diamond have been used to sense rotation~\cite{ajoy_stable_2012,jarmola_demonstration_2021,soshenko_nuclear_2021}, magnetic and electric fields~\cite{taylor_high-sensitivity_2008,Maze_nature2008,dolde_electric-field_2011}, temperature~\cite{kucsko_nanometre-scale_2013,Choi2020} and pressure~\cite{doherty_electronic_2014,Yip2019,Lesik2019,Hsieh2019} with high resolution and precision~\cite{degen_quantum_2017}.  Compatible with industry-standard fabrication process, donor spins in silicon~\cite{morton_solid-state_2008} and defects in silicon carbide~\cite{bourassa_entanglement_2020}, also show potential in building quantum devices such as quantum sensors and quantum memories~\cite{zwanenburg_silicon_2013,morello_donor_2020}. More recently, nuclear spins in rare-earth doped crystals have attracted extensive interest to realize ensemble-based quantum memory and quantum network~\cite{ruskuc_nuclear_2022}.

Scaling up the system size and improving the quantum gate fidelity have been major research directions in quantum information science for realizing quantum advantages. One challenge is the decoherence of the system due to its coupling to a noisy and inhomogeneous environment, as well as the lack of an in-depth characterization of these factors. In particular, when using nuclear spins as memories or sensors, hyperfine-coupled electronic spins are usually needed for the nuclear spin control and readout, but the coherence time becomes limited by the hyperfine interaction. Although traditional dynamical decoupling methods~\cite{suter_colloquium_2016} in the spirit of spin echo~\cite{hahn_spin_1950} protect the quantum system from these environmental noise sources, these sequences typically erase the stored quantum information, and break the sensor's sensitivity to the low-frequency target field. Potential solutions to such a dilemma include controlling the spin bath~\cite{bauch_ultralong_2018,knowles_observing_2014,chen_protecting_2018}, encoding qubits with decoherence-free states~\cite{hart_n_2021,jarmola_demonstration_2021,soshenko_nuclear_2021,reiserer_robust_2016} or utilizing motional narrowing effects~\cite{maurer_room-temperature_2012}, which impose additional complexities and requirements such as fast and high-fidelity control~\cite{chen_protecting_2018} and preparation of double-quantum states~\cite{hart_n_2021,jarmola_demonstration_2021,soshenko_nuclear_2021}. 

\begin{figure*}[htbp]
\centering \includegraphics[width=\textwidth]{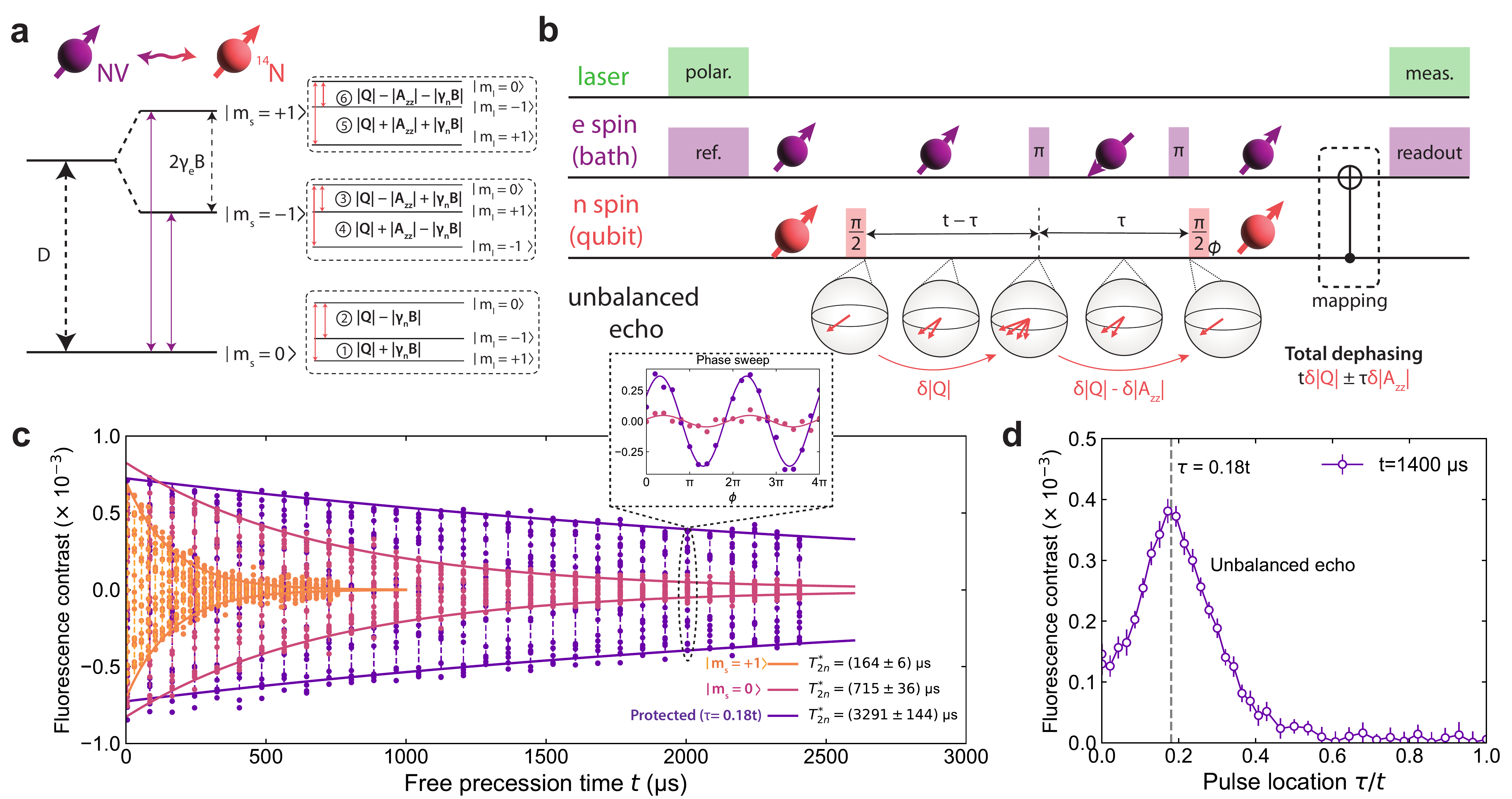}
\caption{\label{Fig_ExperimentalResults} \textbf{Coherence protection schematic and results.} a. Energy levels of NV ground state. b. Coherence protection sequence. An initial laser pulse (followed by a RF $\pi$ pulse not shown in the sequence) polarizes the NV to $\ket{m_S=0,m_I=+1}$ ($\ket{m_S=0,m_I=0}$) state. The nuclear spin state $(\ket{0}+\ket{\pm1})/\sqrt{2}$ is prepared with a $\pi/2$ RF pulses. A spin flip microwave $\pi$ pulse of the NV electronic spin is applied to flip the electronic spin state from $\ket{0}$ to $\ket{\pm1}$. The final electronic spin $\pi$ pulse transfer the NV state back to $\ket{0}$. The $\pi/2$ pulse applied to the nuclear spin associated with a conditional $\pi$ pulse to the electronic spin performs a readout of the nuclear spin population. c. Free evolution (Ramsey) measurements of unprotected and protected nuclear spin states $(\ket{0}+\ket{-1})/\sqrt{2}$. At each time $t$, we sweep the phase of the last $\pi/2$ pulse on the nuclear spin to measure the coherence as shown in the inset. The amplitude of the phase sweep measurements are fit to exponential decay functions. d. Sweep of the pulse location $\tau/t$ with $t=1400\ \mu$s. The fluorescence constrast is measured through similar phase sweeps.
}
\end{figure*}

In this work, we propose a novel \textit{unbalanced echo} method to protect quantum coherence against spatial or temporal variation of the intrinsic interactions, which resembles a typical spin echo process but preserves the quantum information acquired during the qubit's free evolution. We study the protection of the native $^{14}$N nuclear spin near nitrogen-vacancy (NV) center in diamond and show that the optimal protection is achieved by flipping the NV electronic spin only once during the free evolution. Here we exploit hyperfine interaction, which is usually detrimental to the qubit coherence, to cancel out the noise effects. The flip time is set by the relative variation of the quadrupole $Q$ of the nuclear spins and nuclear-NV hyperfine interaction strength $A_{zz}$.  Experimentally, we achieve an order of magnitude improvement of the coherence time of an ensemble of $\sim10^{10}$ nuclear spin qubits. Theoretically, we develop a general first-principles approach to predict the strain and temperature dependence of these intrinsic interactions. Our theoretical results indicate a strong correlation between different interactions dominated by the common phonon coupling, which enable robust coherence protection under extreme temperature or pressure conditions (e.g., $400\times$ coherence improvement under a 25 K temperature variation).  Moreover, our methods can be used to probe temperature and strain distributions in materials with nanoscale resolution. Our results pave the way to building highly sensitive sensors such as gyroscopes, thermometers and strain sensors, as well as long-lived and highly stable quantum memories using an ensemble of solid-state spins. 

\section*{Results}
\subsection*{Coherence protection with unbalanced echo}
The NV center is an atom-like spin defect in diamond consisting of a vacancy adjacent to a nitrogen substitute with a spin-1 electronic ground state which can be optically polarized to $\ket{m_S=0}$ and readout in ambient condition~\cite{doherty_nitrogen-vacancy_2013}, making it a powerful platform for quantum sensors~\cite{taylor_high-sensitivity_2008,Maze_nature2008,kucsko_nanometre-scale_2013,Choi2020,Li2022,Rendler_nc2017}, networks~\cite{PompiliScience2021,Humphreys_nature2018}, and memories~\cite{reiserer_robust_2016}. The nitrogen nuclide is also a spin-1 that can be polarized and read out via its hyperfine interaction with the electronic spin~\cite{fischer_optical_2013,chakraborty_polarizing_2017}. The long coherence time of the nuclear spin enables applications such as gyroscopes~\cite{ajoy_stable_2012,jaskula_cross-sensor_2019,jarmola_demonstration_2021,soshenko_nuclear_2021} and quantum registers~\cite{dreau_single-shot_2013,neumann_single-shot_2010,Liu_single_shotPRL2017}.

\begin{figure}[htbp]
\centering \includegraphics[width=0.49\textwidth]{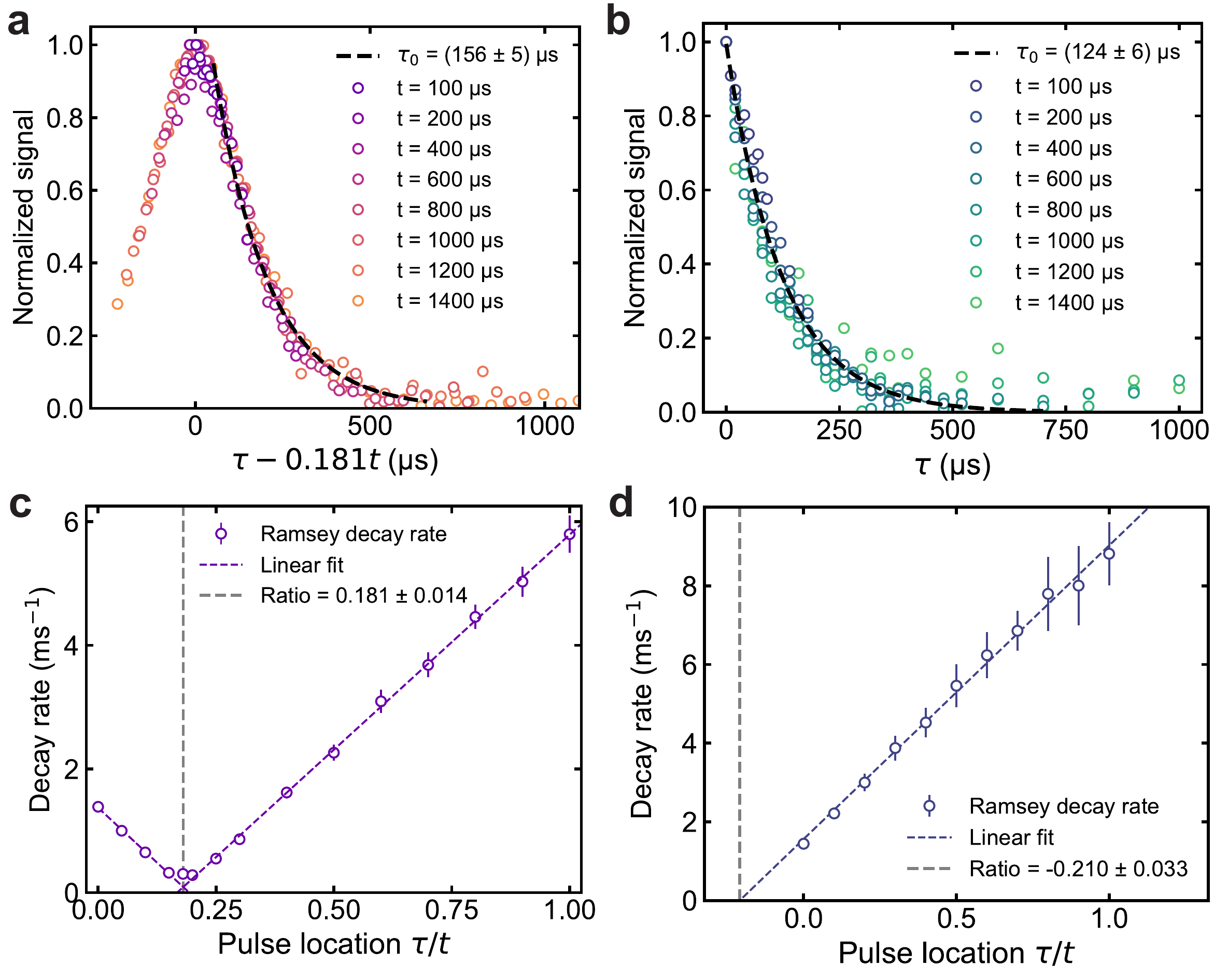}
\caption{\label{Fig_ExperimentalResults_Fig2} \textbf{Characterization of the inhomogeneity.} a. Sweep of pulse location $\tau/t$ under different free precession times $t$. The nuclear spin state is prepared to $(\ket{0}+\ket{-1})/\sqrt{2}$ and the electronic spin flip is between $\ket{m_S=0}$ and $\ket{m_S=+1}$. b. Sweep of pulse location $\tau/t$ with the nuclear spin state prepared to $(\ket{0}+\ket{+1})/\sqrt{2}$ and the electronic spin flip is between $\ket{m_S=0}$ and $\ket{m_S=+1}$. c. Decay rate ($1/T_{2n}^*$) measurement as a function of the pulse location $\tau/t$. The nuclear spin state is prepared to $(\ket{0}+\ket{-1})/\sqrt{2}$ and the electronic spin flip is between $\ket{m_S=0}$ and $\ket{m_S=+1}$. d. Same experiment as c except for the nuclear spin initial state $(\ket{0}+\ket{+1})/\sqrt{2}$.}
\end{figure}

When focusing on the nuclear spin $I$ of the NV center aligned in a magnetic field $B$, the Hamiltonian includes quadrupole and hyperfine interaction terms with the coupling between an electronic spin and a magnetic field such that
\begin{equation}
    H_n=QI_z^2+m_SA_{zz}I_z+\gamma_nBI_z,
\end{equation}
where $I_z$ is the spin-1 $z$ operator, and we assume the NV electronic spin is in its eigenstate $\ket{m_S}$.
As shown in Fig.~\ref{Fig_ExperimentalResults}a, the single quantum (SQ) nuclear spin transitions between $\ket{m_I=0}$ and $\ket{m_I=\pm1}$ have six different frequencies $\omega_{1,2}=|Q|\pm|\gamma_nB|$, $\omega_{3,4}=|Q|\mp|A_{zz}|\pm|\gamma_nB|$ and $\omega_{5,6}=|Q|\pm|A_{zz}|\pm|\gamma_nB|$ corresponding to the electronic spin states $\ket{m_S=0}$, $\ket{m_S=-1}$ and $\ket{m_S=+1}$, respectively. 

In many scenarios, quantum information is stored in a superposition state undergoing free evolution $\ket{\psi(t)}=(\ket{0}+e^{i\varphi}\ket{-1})/\sqrt{2}$, which accumulates a quantum phase $\varphi=\omega t$. When such a phase is used for sensing the qubit frequency $\omega$, one measures the overlap between the final state and initial state yielding a Ramsey oscillation signal $S(t)=[1+\cos(\omega t)]/2$. For an ensemble of qubits with a noisy environment, the measured signal is an average quantity $S(t)=[1+\langle\cos(\omega t)\rangle]/2$ over the frequency distribution $\omega=\omega_0+\delta\omega$. When $\delta\omega$ satisfies a zero-mean Lorenzian distribution with a half-width $\sigma=1/T_{2}^*$, the signal has an exponential decay envelope $\langle\cos(\omega t)\rangle=\cos(\omega_0t)e^{-t/T_2^*}$ characterizing the loss of the quantum information. When the NV is in a state $\ket{m_S=0}$, the variation of the SQ transition frequency $\delta\omega=\delta|Q|+\gamma_n\delta|B|$; when the NV electronic spin is in a state $\ket{m_S=\pm1}$, the variation is $\delta\omega=\delta|Q|\mp\delta|A_{zz}|+\gamma_n\delta|B|$. 
The variation of the magnetic field $\delta B$ is usually minimized when designing the experimental setup, and the small gyromagnetic ratio $\gamma_n/\gamma_e\sim10^{-4}$ introduces negligible coupling to the magnetic noise. 
Thus, the coherence time of the nuclear spin ensemble is largely limited by the variations in quadrupole and hyperfine interactions. The most common mechanisms include temperature and strain inhomogeneities introduced by laser heating and natural or applied strain in the material. 

A powerful strategy for coherence protection is based on an echo sequence with a spin flip pulse applied in the middle of the free evolution such that the phases accumulated during the two halves of the total evolution cancel each other. However, this process also cancels the useful quantum phase information $\omega_0$.

In contrast, we apply a spin flip pulse to the NV electronic spin. This is equivalent to effectively inverting the bath/environment state, as the NV mediates the effects of temperature and strain. The protocol works as long as the noise source affects both the quadrupole and hyperfine interactions regardless of the mechanism governing the relationship between the noise and the interactions.
For example, in optically addressed solid-state spin ensembles, the laser heating introduces temperature variations $\delta{T}$. 
Thus, the variation of both interactions are correlated and can be expressed as $\delta Q=\alpha_Q \delta{T}$, $\delta A_{zz}=\alpha_A \delta{T}$. To protect the target nuclear spin state $\ket{\psi(0)}$ prepared under the NV state $\ket{m_S=0}$, we flip the NV to $\ket{m_S=+1}$ at time $t-\tau$ in a free evolution duration $t$ (Fig.~\ref{Fig_ExperimentalResults}b). The total phase accumulation of the target nuclear spin under a temperature variation $\delta{T}$ is then $\varphi(\delta{T})=\varphi_0+\delta\varphi$ with
\begin{equation} 
    \varphi_0=-tQ+\tau A_{zz}+\gamma_nBt,\quad \delta\varphi=-(t\alpha_Q-\tau\alpha_A)\delta{T}.
    \label{eq:Phase}
\end{equation}
When the pulse location $\tau/t$ is set to the ratio $\alpha_Q/\alpha_A$, the noise $\delta\varphi$ will be completely canceled while the static part $\varphi_0$ still remains sensitive to the magnetic field (or a rotation phase in gyroscopes), the mean quadrupole and hyperfine interactions, which are useful for various sensing applications.

\begin{figure}[htbp]
\centering \includegraphics[width=0.495\textwidth]{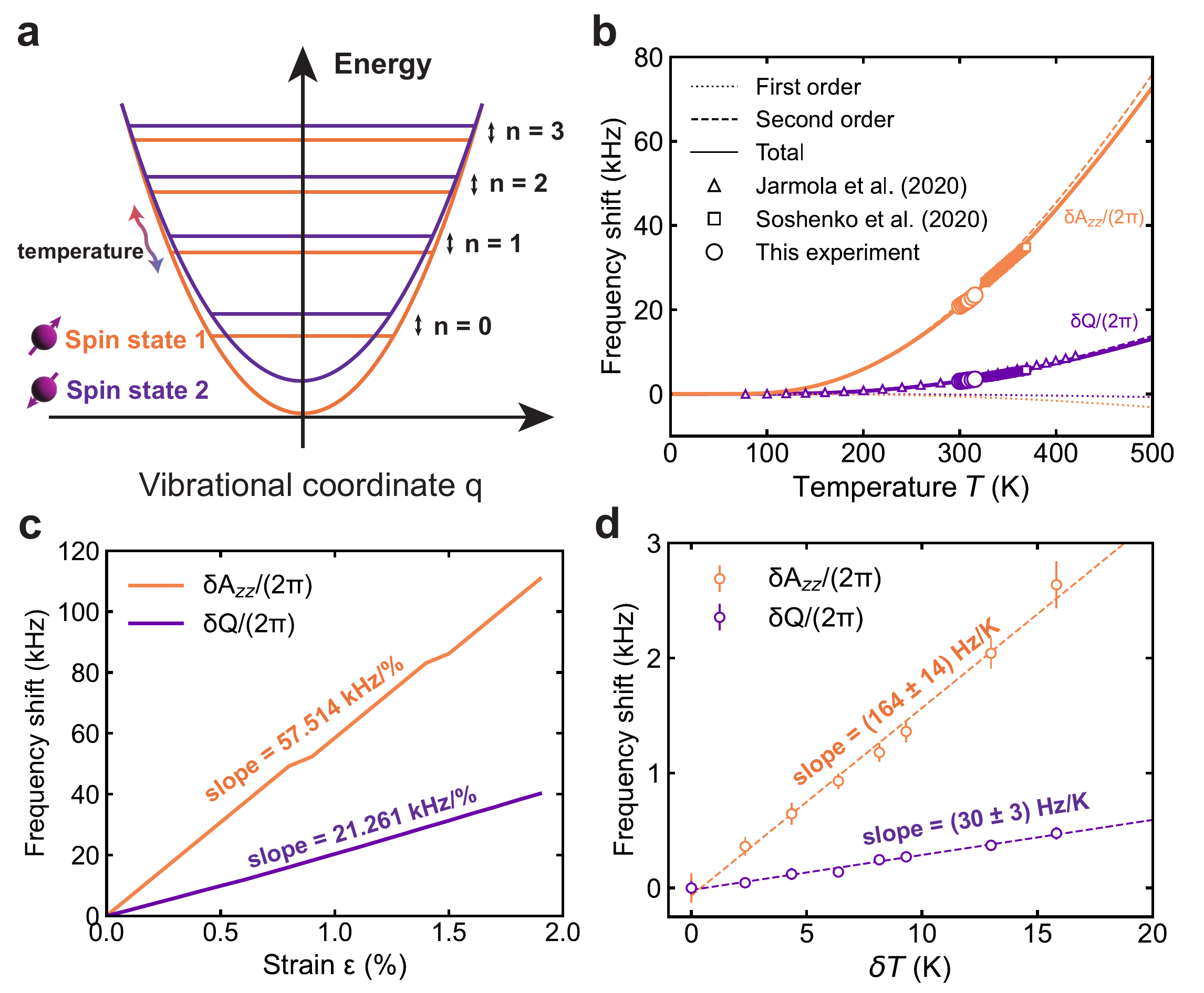}
\caption{\label{Fig_Theory1} \textbf{Temperature and strain dependencies.} a. Schematic showing the energy levels of the spin coupled to the phonon. b. Calculated frequency shifts of the quadrupole ($\delta Q$) and hyperfine ($\delta A_{zz}$) interactions as a function of the temperature. Dashed and solid lines represent the first-order, second-order and total shifts predicted by the theory. Our experimental measurements, as well as data extracted from Refs.~\cite{soshenko_temperature_2020,jarmola_robust_2020} are also plotted. Since the offset of the frequency is a free parameter, experimental results are shifted to match the simulation results. c. Calculated frequency shifts of the quadrupole ($\delta Q$) and hyperfine ($\delta A_{zz}$) interactions as a function of the applied strain. The slopes of two curves are obtained by fitting the simulation results to $y=c_0+c_1x$. The dependence of hydrostatic pressure can be calculated using diamond's bulk modulus $K=443$~GPa such that $P=-K\delta V/V=-3K\epsilon$, yielding $\delta A_{zz}/\delta P=(2\pi)4.33$~kHz/GPa and $\delta Q/\delta P=(2\pi)1.60$~kHz/GPa. d. Experimental measurements of frequency shifts as a function of temperature shifts.
}
\end{figure}

Our protocol is experimentally demonstrated in Fig.~\ref{Fig_ExperimentalResults}c, where we compare the nuclear spin coherence times with and without protection sequences, showing one order of magnitude improvement in $T_{2n}^*$. We note that for both unprotected cases, the slightly longer coherence time under $\ket{m_S=0}$ is caused by the vanishing hyperfine interaction contribution in the transition frequency. To visualize the refocusing effect, we fix the free precession time and sweep the spin flip location in Fig.~\ref{Fig_ExperimentalResults}d, where an \textit{unbalanced echo} signal is observed at $\tau/t=0.18$.

\subsection*{Characterization of temperature and strain distribution}
Our method is also a probe of the material-dependent ratio $\alpha_Q/\alpha_A$, and can be used to characterize the temperature distribution in the material by analyzing the shape and decay constant in the unbalanced echo and free evolution measurements. For instance, for a Lorenztian distribution of the temperature with half-width $\sigma_{T}$, the decoherence factor is analytically obtained as
\begin{equation}
    e^{-t/T_{2n}^*}=e^{-|t\alpha_Q\pm \tau\alpha_A|\sigma_{T}}=e^{-|\tau\pm\frac{\alpha_Q}{\alpha_A}t||\alpha_A|\sigma_{T}}
    \label{Eq:decay}
\end{equation}
for nuclear spin states $(\ket{0}+\ket{\pm1})/\sqrt{2}$, respectively~\footnote{In general, the decay is related to the p.d.f. characteristic functions of the temperature and strain variations.}.
In Figs.~\ref{Fig_ExperimentalResults_Fig2}a,b, we measure a series of pulse location sweep experiments under different free precession times. The overlap of the normalized data plotted as a function of shifted time validates the prediction in Eq.~\eqref{Eq:decay} and the exponential shape in the $t$ dependence indicates a Lorenzian distribution of the temperature variation. In Figs.~\ref{Fig_ExperimentalResults_Fig2}c,d, we measure the free precession decay rate $1/T_{2n}^*$ as a function of the pulse location. The linear dependence function validates the Lorenzian model and gives a better estimation of the ratio $\alpha_Q/\alpha_A\approx0.18$, which is consistent with reported values under a similar temperature range with $\delta Q/\delta{T}\approx39\text{ Hz}/\text{K}$ and $\delta A_{zz}/\delta{T}\approx204\text{ Hz}/\text{K}$~\cite{soshenko_temperature_2020,jarmola_robust_2020} yielding $\alpha_Q/\alpha_A\approx 0.19$. The measured coherence times indicate a $\sigma_{T}\sim5$~K temperature inhomogeneity in our sample.

In a general situation when more than one independent noise sources exist, (e.g., temperature and strain), variation of the interaction constants is modified to $\delta Q=\alpha_Q \delta{T}+\beta_Q \delta\epsilon+\cdots$, $\delta A_{zz}=\alpha_A \delta{T}+\beta_A\delta\epsilon+\cdots$. Since the temperature itself will also introduce a strain through thermal expansion (``stress-free strain"), here $\epsilon$ represents strain additionally induced by finite stresses. In these cases, our method can characterize the distribution of different sources given that the material-dependent constants $\alpha,\beta,\cdots$ are separately measured by spin resonance experiments and the ratios $\alpha_Q/\alpha_A$, $\beta_Q/\beta_A, \cdots$ are different. For example, when all noise sources satisfy a Lorentzian distribution with half-widths $\sigma_{T},\sigma_\epsilon,\cdots$, the decoherence factor is then $e^{-\chi}=e^{-|t\alpha_Q\pm \tau\alpha_A|\sigma_{T}-|t\beta_Q\pm \tau\beta_A|\sigma_\epsilon-\cdots}$. 
Thus, by experimentally measuring the qubit decoherence as a function of $t,T$ and comparing the results to the predicted formula, the distribution of the noise sources in the target material can be characterized including both the standard deviations $\sigma_{T},\sigma_\epsilon,\cdots$ and the characteristic functions.

\subsection*{Refocus strong noise variations}

For most cases with weak noise sources such that the interaction variation depends linearly on the variation of a dominant noise source, the ratio $\alpha_Q/\alpha_A$ or $\beta_Q/\beta_A$ remains fixed and the optimal coherence protection is achieved by setting the spin flip location $\tau/t$ to such a fixed ratio. To extend our method to a general condition such as large temperature or strain variations, one needs to evaluate the correlation between quadrupole and hyperfine variations with a more in-depth theoretical understanding of the underlining physical mechanism. The temperature influences the energy level through both (first-order) thermal expansion similar to the pressure-induced strain~\cite{doherty_electronic_2014} and (second-order) phonon distribution~\cite{doherty_temperature_2014,jarmola_robust_2020}, which changes the electron spin density and electric field gradient. The energy diagram forms a spin-dependent ladder separated by the phonon mode frequency as shown in Fig.~\ref{Fig_Theory1}a. The spin transition frequency is then a weighted average of all the phonon states satisfying the Bose-Einstein distribution. Based on this analysis, we develop a predictive first-principles method to calculate the quadrupole and hyperfine variations. Our method is also generic to the calculation of other constants such as zero-phonon line, zero-field splitting in solid-state spin defects~\cite{Tang_Temperature_2022}.

In Fig.~\ref{Fig_Theory1}c, we calculate the hydrostatic strain dependence of both interaction variations $\delta Q$, $\delta A_{zz}$ and predict a relatively large linear range up to a $\sim 26.6$~GPa pressure (correspond to $\epsilon=2\%$), which has been validated in a recent experiment~\cite{arxiv2203.10511}. In high-pressure spectroscopy measurements, the strain inhomogeneity of the material could reach GPa levels resulting in limited spectral resolution~\cite{wang_ac_2021,arxiv2203.10511}. The predicted large linear range shows that our method can perfectly refocus at least a $\sim 26$~GPa pressure inhomogeneity, which paves the way to characterizing quantum materials in extreme conditions using quantum sensors. In addition, the strain (or electric field) has been used to tune the transitions in color centers in diamond~\cite{udvarhelyi_spin-strain_2018-1,meesala_strain_2018}, as well as the hyperfine~\cite{mansir_linear_2018,wolfowicz_conditional_2014} and quadrupole interactions~\cite{franke_interaction_2015} in the donor nuclear spins in silicon. Our results make it possible to design high-fidelity strain control protocols for these applications.

\begin{figure}[htbp]
\centering \includegraphics[width=0.495\textwidth]{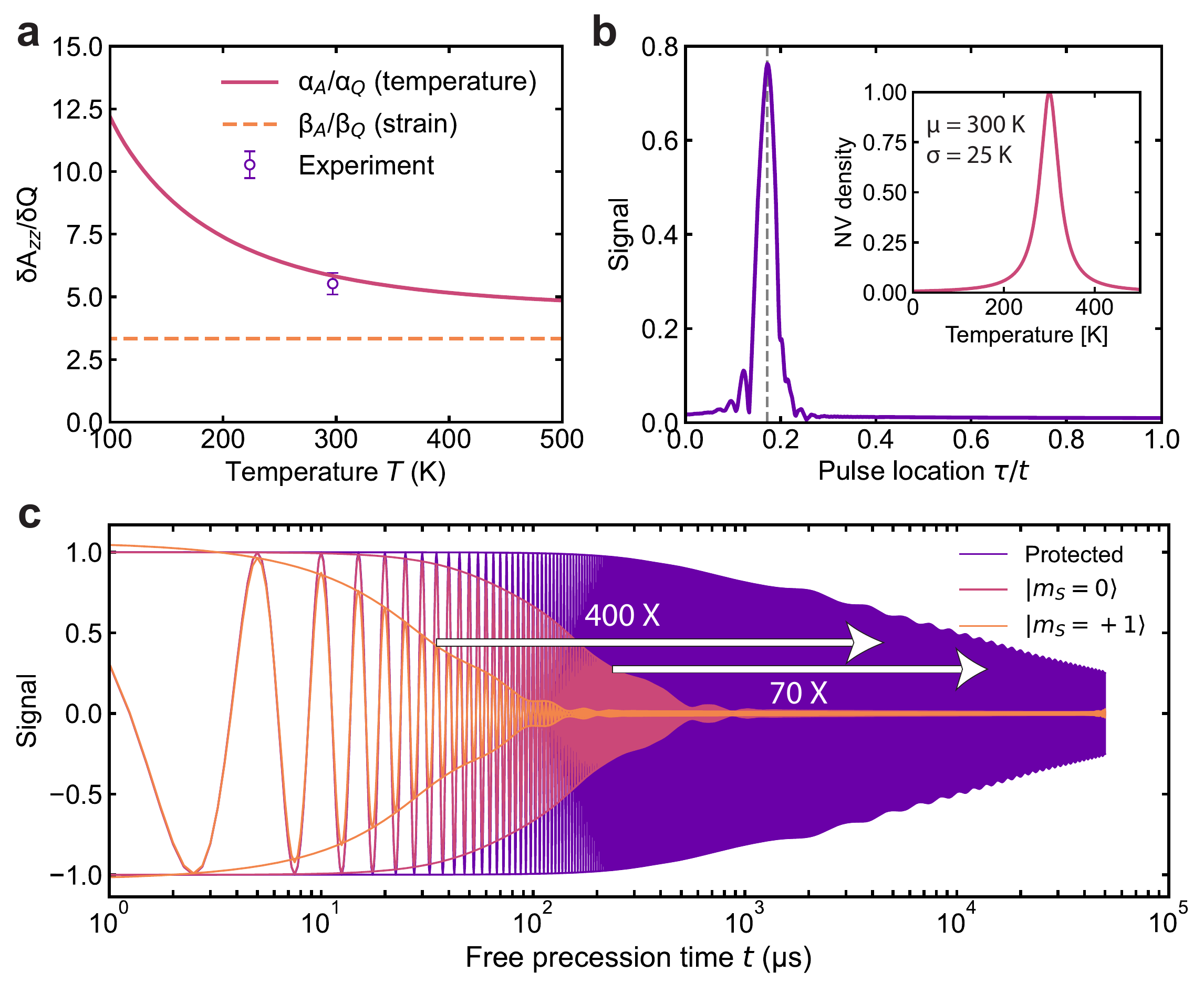}
\caption{\label{Fig_Simulation1} \textbf{Coherence protection under a large temperature inhomogeneity.} a. Calculated ratios $\alpha_A/\alpha_Q$ and $\beta_A/\beta_Q$ as a function of temperature. The ``strain" data plots the ratio between the two slopes in Fig.~\ref{Fig_Theory1}c. b. Pulse location sweep with $t=2$~ms for a qubit ensemble with a temperature inhomogeneity satisfying a Lorentzian function with $(\mu,\sigma)=$(300, 25)~K (inset). The nuclear spin state is chosen to be $(\ket{0}+\ket{-1})/\sqrt{2}$ and the electronic spin flip is between $\ket{m_S=0}$ and $\ket{m_S=+1}$.  c. Simulation of unprotected Ramsey experiments for an initial nuclear spin state $(\ket{0}+\ket{-1})/\sqrt{2}$ under electronic spin states $\ket{m_S=+1}$ (orange), $\ket{m_S=0}$ (pink), as well as protected nuclear spin Ramsey experiment (purple) with spin flip time $\tau/t=0.172$ obtained in the unbalanced echo simulation shown in b. }
\end{figure}

In Fig.~\ref{Fig_Theory1}b,  the temperature dependence of both the quadrupole and hyperfine interactions show a nonlinear shape. The match between the experiment and calculation validates our theory and reveals the dominant role of phonons in determining the temperature dependence of NV energy levels~\cite{Tang_Temperature_2022}. Fig.~\ref{Fig_Simulation1}a shows that the ratios between the two variations for both temperature and strain noise sources satisfy $\alpha_A/\alpha_Q,\beta_A/\beta_Q>1$ (required by $t/\tau>1$ in unbalanced echo sequences) in broad temperature and strain ranges, which demonstrates the broad applicability of our protection method. In addition, since the temperature and strain have different ratios $\alpha_A/\alpha_Q\neq\beta_A/\beta_Q$, it is possible to characterize and distinguish the temperature and strain distributions in the material. Specifically, the ratio between the hyperfine and qudrupole variations obtained from our unbalanced echo experimental result matches the theoretically predicted ratio due to the temperature variation (Fig.~\ref{Fig_Simulation1}a), which indicates that the noise source in our sample is dominated by temperature variations.


Ideally, the protection method can completely refocus the noise only when the variations of the quadrupole and hyperfine interactions have a constant ratio (e.g. in a small range). However in a large variation range where the two interactions do not have an exact constant ratio (Fig.~\ref{Fig_Simulation1}a) but have similar shapes (Fig.~\ref{Fig_Theory1}b), it is still possible to achieve a relatively good protection.
Based on the first-principles results, we numerically study the coherence protection of a large temperature inhomogeneity  as shown in Fig.~\ref{Fig_Simulation1}b with the optimal pulse location $\tau/t$ obtained by simulating the unbalanced echo experiment. In Fig.~\ref{Fig_Simulation1}c, we compare the coherence times under both the unprotected and protected sequences. For a large temperature inhomogeneity $\sigma_{T}=25$~K, more than $10$~ms coherence time can be achieved (assuming no other noise sources), which shows a 400-fold improvement. 

\section*{Discussions and conclusions}

In our practical experiments, the optimal coherence time under protection of $\approx 3.5$~ms (under $\tau/t=0.2$) is ultimately limited by other noise sources such as field inhomogeneities and magnetic noise induced by other electron or nuclear spins in diamond~\cite{SOM}. A similar process canceling inhomogeneity in quadrupole interaction while preserving sensitivity to the static field is to use the double-quantum (DQ) superposition state $(\ket{+1}+\ket{-1})/\sqrt{2}$ under the electronic spin states $\ket{m_S=0}$~\cite{jarmola_demonstration_2021,soshenko_nuclear_2021,hart_n_2021}, which gives a coherence time $T_{2,DQ}^*=3.9\pm0.3$~ms. In comparison to the DQ method which requires complicated control for state preparation and readout, our method only requires a simple SQ control and therefore, it is more robust to pulse errors. Our method also preserves the information of quadrupole and hyperfine interactions useful for quantum sensing [Eq.~\eqref{eq:Phase}]. 
Moreover, our method is applicable to other spin baths. For example, for a spin-1/2 electronic bath, our method cancels the noise variation when satisfying $t\delta Q+(t-2\tau)\delta A=0$, while the DQ fails due to the hyperfine interaction contribution in the free evolution.

Our method can completely refocus the spatial inhomogeneity dominated by a common source or several common sources with the same ratio between the variations of two interactions. When more than one effect with different ratios dominate, the optimal performance highly depends on those material constants $\alpha,\beta,\cdots$. However, one can still characterize the relative contributions of different sources and optimize the quantum devices on purpose based on the obtained information, for example based on finite-element calculations of the thermomechanical fields ${T}({\bf x})$, $\mathcal{\epsilon}({\bf x})$ where $\bf x$ is the spatial coordinate of the spin defect.

We note that the same method can also decouple time varying noise sources for both single and ensemble spins. In that scenario, it is convenient to keep the unbalanced echo time shorter than the noise correlation time and repeat the sequence multiple times, similar to the strategy adopted in multi-pulse dynamical decoupling methods~\cite{suter_colloquium_2016}. 

In conclusion, we propose to use unbalanced echo to protect the solid-state spin qubit by controlling the hyperfine-coupled spin bath where the variations of the quadrupole and hyperfine interactions are canceled with each other. The same sequence can be used to characterize the temperature and strain distributions in the target material. To validate the broad applicability of our method, we develop a predictive first-principles approach to calculate the temperature and strain dependence of different interactions in generic solid state spins and reveal the mechanism of the correlation between these variations due to phonon coupling. Our theoretical results along with numerical calculations show the possibility of completely refocusing $>20$~GPa pressure inhomogeneity and achieving a $400\times$ coherence improvement even with tens of Kelvin temperature variations. 

A few promising directions include but are not limited to: combining our method with existing dynamical decoupling methods to explore the sensing of non-static quantities; concatenate our sequence with other coherence protection methods such as driving other bath spins to further push the coherence limit; 
develop nanoscale sensors to probe temperature, strain and electric field distributions with high spectral resolution under extreme conditions such as high pressure; develop high-fidelity strain control with high spin in silicon donors; build robust and long-lived quantum memories and registers.


\section*{Methods}
\textbf{NV ground state Hamiltonian and diamond sample.}
The spin Hamiltonian of an NV center aligned in a magnetic field $B$ is
\begin{equation}
    H=DS_z^2+\gamma_eBS_z+QI_z^2+\gamma_nBI_z+A_{zz}S_zI_z
\end{equation}
where $S_z$, $I_z$ denote the operators for NV electronic spin and nuclear spin, $D\approx(2\pi)2.87$~GHz is the zero-field splitting (ZFS), $Q\approx-(2\pi)4.945$~MHz the nuclear-spin quadrupole term, $A_{zz}\approx-(2\pi)2.16$~MHz the hyperfine interaction between the nuclear spin and electronic spin. $\gamma_e=(2\pi)2.8025$~MHz/G, $\gamma_n=-(2\pi)0.3077$~kHz/G are the corresponding gyromagnetic ratios characterizing the coupling strength between the spin and magnetic field.
We note that here the transverse hyperfine interaction term $A_\perp(S_xI_x+S_yI_y)$ is neglected under a secular approximation when $A_\perp,\gamma_eB\ll D$~\cite{jarmola_demonstration_2021}. 

Our diamond sample (2~mm$\times$2~mm$\times$0.5~mm) is the same one as in Ref.~\cite{jaskula_cross-sensor_2019} grown using chemical-vapor deposition (CVD) by Element Six on an electronic-grade diamond substrate. The top-surface NV layer is isotopically purified with 99.999\% $^{12}$C with 20 ppm $^{14}$N concentration. By focusing a 0.4~W 532 nm green laser beam to a $\sim30~\mu$m spot, we simultaneously address about $10^9\sim10^{10}$ spins.
The diamond is mounted on a piece of home-designed printed circuit board (PCB) with a 0.7 mm diameter loop to apply the microwave and radio-frequency field. The fluorescence coming out of the sides of the diamond is collected by three photodiodes (Hamamatsu S8729) attached to the PCB board. The compact design is flexible to be integrated with other instruments in different applications~\cite{SOM}. A pair of permanent magnets apply a magnetic field $B=239$~Gauss along the NV axis, which polarizes the nuclear spin state to 70\% in $\ket{m_I=+1}$.


\textbf{First-principles method.}
In recent years, theoretical and experimental studies over the dependence of the zero-phonon-line (ZPL) and ZFS~\cite{acosta_temperature_2010,doherty_temperature_2014}, as well as nuclear quadrupole~\cite{jarmola_robust_2020} and hyperfine~\cite{soshenko_temperature_2020,barson_temperature_2019} interactions in NV centers have been reported. The temperature influences the energy level through thermal expansion that changes the equilibrium lattice displacement similar to the pressure-induced strain~\cite{doherty_electronic_2014}, as well as the dynamic phonons that couple to the energy level in a second order manner~\cite{doherty_temperature_2014,jarmola_robust_2020}. Previous first-principles calculations only provided a validation of the first order effect while the explanation of the second order effect remains phenomenological~\cite{doherty_temperature_2014,ivady_pressure_2014}, so recently we extended the computational toolkit to the second order and provided a parameter-free prediction of the complete energy levels~\cite{Tang_Temperature_2022}. In this work we focus on the physical mechanism related to the quadrupole and hyperfine interactions, while a more complete computational methodology and theoretical analysis including the ZPL, ZFS for general color centers will be published separately~\cite{Tang_Temperature_2022}.

The phonons couple to the energy levels of hyperfine and quadrupole interactions by changing the spatial distribution of the electron spin density and electric field gradient when varying the temperature. Similar to harmonic oscillators, the energy levels of a phonon form a nearly equidistant quantum energy ladder separated by the phonon mode frequency as shown in Fig.~\ref{Fig_Theory1}a. 
We adopt a quasiharmonic approximation in which the vibrational energy level spacings at higher quantum numbers is ignored, so that the vibrational modes are temperature independent and remain harmonic. The anharmonic displacement of the average nuclei positions at higher vibrational quantum numbers is then approximated by the thermal expansion induced static strain, so that the interaction frequency $\nu=Q,A_{zz}$ we are interested can be expressed in terms of the phonon coordinate $q$ as
\begin{equation}
\nu\left(T\right) \simeq \nu_{0}+\left. \frac{\partial \nu}{\partial q_{\rm{ex}}}\right|_{q_{i}=0}q_{e x}(T)+\left.\sum_{i} \frac{1}{2} \frac{\partial^{2} \nu}{\partial q_{i}^{2}}\right|_{q_{i}=0}\left\langle q_{i}^{2}\right\rangle  \label{Eq:theory}
\end{equation}
where $i$ denotes different phonon modes and $q_{\rm{ex}} (T)$ denotes temperature dependent displacement of the average nuclei position along the normal
displacement coordinate from low temperature equilibrium. The normal displacement coordinate $q_{\rm{ex}}$ corresponds to the `breathing' vibrational mode of the whole lattice, and is both isotropic and homogeneous. The temperature influences the frequency through both an electron configuration change which induces the change of the derivative, as well as an average population change in different phonon levels $n$ satisfying the Bose-Einstein distribution. 
The strain influences the equilibrium atom configuration in a similar way as the first order term in Eq.~\eqref{Eq:theory}. 

In the density functional theory (DFT) calculation, we use a $4\times 4\times 4$ rhombohedral supercell with a single NV in the center. We use the  Vienna ab-initio simulation package (VASP) with a cut-off energy of 520 eV~\cite{kresse1996efficient,kresse1999ultrasoft} and employ the projector-augmented-wave (PAW) method to calculate the electronic structure with the fully relaxed atomic configuration at zero temperature. Then we perform spin-unrestricted calculations using the generalized gradient approximation (GGA) with the Perdew-Burke-Ernzerhof (PBE) functional for electron exchange-correlation~\cite{perdew1996generalized}. Please check the complete details including the phonon calculation published in our separate work in Ref.~\cite{Tang_Temperature_2022}.

\section*{Acknowledgement}
This work were supported in part by DARPA DRINQS program (Cooperative Agreement No. D18AC00024), HRI-US, NSF DMR-1923976, NSF DMR-1923929 and NSF CMMI-1922206. A.R.B. acknowledges support from a National Science Foundation Graduate Research Fellowship under Grant No. DGE-174530. The calculations in this work were performed in part on the Texas Advanced Computing Center (TACC) and the MIT engaging cluster. G.W. thanks Thanh Nguyen for help in figure revision.

\section*{Author contributions}
G.W. and P.C. proposed the method and designed the experiment. G.W. performed the experiment and data analysis. A.R.B., H.T., J.L. developed the computational methods and A.R.B., H.T. implemented the first-principles calculations. M.C., C.L. contributed to part of the data analysis. G.W., P.C. and A.R.B., H.T., H.X., J.L. contributed to the theoretical study. G.W. performed the numerical simulation of the coherence protection. All authors discussed the results.

\section*{Competing financial interest}
The authors declare no competing financial interests.


\clearpage
\setcounter{section}{0}
\setcounter{equation}{0}
\setcounter{figure}{0}
\setcounter{table}{0}
\setcounter{page}{1}
\makeatletter
\renewcommand{\theequation}{S\arabic{equation}}
\renewcommand{\thesection}{S\arabic{section}}
\renewcommand{\thefigure}{S\arabic{figure}}


\title{Supplementary Information}
\maketitle
\begin{widetext}

\section{Experimental setup}
\begin{figure*}[htbp]
\centering \includegraphics[width=0.7\textwidth]{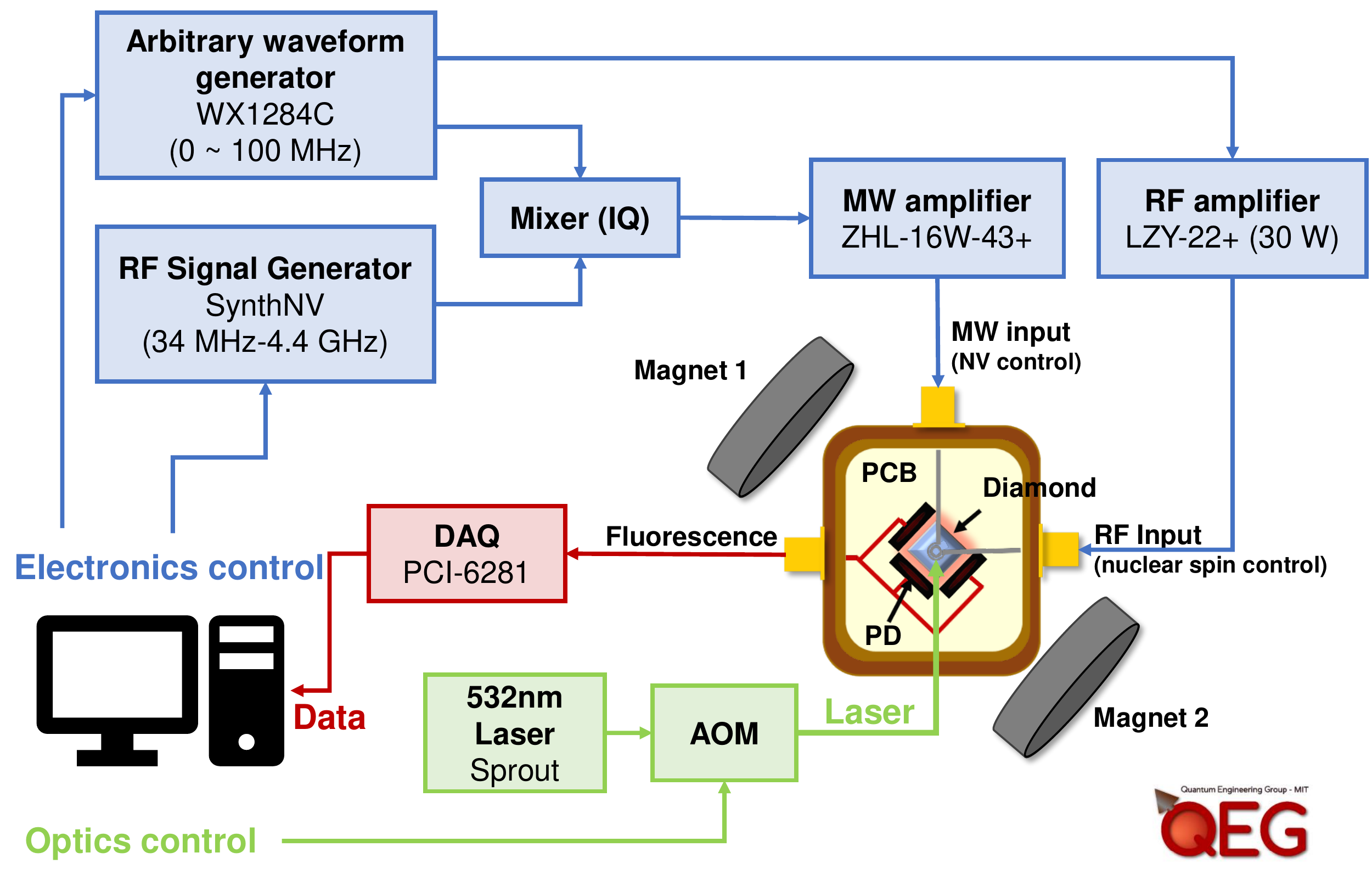}
\caption{\label{Supp_Fig_Setup} \textbf{Experimental setup.}}
\end{figure*}
The schematic of our experimental setup is shown in Fig.~\ref{Supp_Fig_Setup}. The diamond sample is mounted on a piece of home-designed printed circuit board (PCB) with a 0.7 mm diameter loop to apply the microwave and radio-frequency field. A pair of permanent magnets create a magnetic field along the NV axis to split the degeneracy between $\ket{m_S=\pm1}$.
An arbitrary waveform generator (AWG) generate microwave and radio-frequency fields to control the NV electronic spin and nuclear spin. Three synchronized channels are used: two of them (I and Q) generate a 20~MHz frequency with a 90\degree phase shift to mix with a GHz-level microwave signal generated by a fixed microwave source for the control of the NV electronic spin; the third channel generates a radio-frequency in the range of $2\sim8$~MHz to control the nuclear spin. The signals are amplified by two different amplifiers and then input to the PCB board. The radio frequency is directly generated by the AWG (one channel).
The fluorescence coming out of the sides of the diamond is collected by three photodiodes attached to the PCB board.
A pulse blaster (PulseBlasterESR-PRO-500) is used to generate synchronized triggers to control the on-off time of electronics, laser and data acquisition.

\section{Measurement details}
\textit{Nuclear spin readout.} The NV electronic spin state population in $\ket{m_S=0}$ is directly read out from the fluorescence intensity. The readout of the nuclear spin of the $^{14}$N is through its mapping to the NV electronic spin state. For example, to measure the population in $\ket{m_I=0}$ for a nuclear spin state $\alpha\ket{m_I=0}+\beta\ket{m_I=+1}$ with the NV center in $\ket{m_S=0}$, we apply a spin-selective $\pi$ pulse to transfer the population of the NV from $\ket{m_S=0}$ to $\ket{m_S=+1}$ conditional on the nuclear spin state $\ket{m_I=+1}$ by tuning the pulse frequency to the resonance frequency between $\ket{m_S=0,m_I=+1}$ and  $\ket{m_S=+1,m_I=+1}$ ($D+\gamma_eB+A_{zz}m_I$). In this process, the state $\ket{m_S=0}\otimes (\alpha\ket{m_I=0}+\beta\ket{m_I=+1})$ is transferred to  $(\alpha\ket{m_S=0}\otimes \ket{m_I=0}+\beta\ket{m_S=+1}\otimes\ket{m_I=+1})$ such that the nuclear spin state is entangled with the electronic spin state. Then the measurement of the population in $\ket{m_S=0}$ is equivalently a measurement of the nuclear spin population in $\ket{m_I=+1}$. Similar methods are used in reading other nuclear spin states. 

In all of the experiments in this work, we calibrate and use $\sim 500$~ns pulse length (1~MHz Rabi power) for the conditional electronic spin flip and $\sim80$~ns pulse length (6.25~MHz Rabi power) for the unconditional electronic spin flip. For the conditional electronic spin flip, we tune the pulse frequency to the resonance frequency of the corresponding nuclear spin hyperfine level, while for the unconditional electronic spin flip, we tune the frequency to the middle of the two resonance frequencies to make it symmetric to both states. For example, when the nuclear spin states $\ket{m_I=0},\ket{m_I=+1}$ are used, we tune the pulse frequency of the unconditional electronic spin flip to the average value of the two hyperfine transition frequencies between $\ket{m_S=0}$ and $\ket{m_S=+1}$ ($D+\gamma_e B+A_{zz}/2$). 

\textit{Phase sweep measurements.} In this work, we measure the coherence of the nuclear spin by sweeping the phase $\phi$ of the last $\pi/2$ pulse, which measures the final state's population in  $\ket{\psi_f}=(\ket{0}+e^{i\phi}\ket{\pm1})/\sqrt{2}$ in the rotating frame, yielding a signal $S(t)=[1+e^{-t/T_2}\cos(\phi+\phi_0)]/2$, where $\phi_0=\delta\omega t$ is a constant set by the frequency detuning of the pulses ($\delta\omega\approx0$ in this work). When taking into account the practical experiment where the signal is calculated by normalizing the fluorescence with a reference in the beginning of the sequence (main text Fig.~1), the signal formula is then modified to 
\begin{equation}
    S(t)=c_0-\frac12 c+\frac12 c\cos(\phi+\phi_0)e^{-t/T_2},
\end{equation}
where $c$ is the signal contrast and $c_0\approx 1$ is a constant shift due to the sequence wait time (degrading the fluorescence due to $T_1$ relaxation process of the NV center). 
Thus, by performing a phase sweep, the useful signal contrast $c$ reflecting the qubit coherence is extracted by fitting to the cosine function (shown in the main text Fig.~1c inset).

\section{Coherence time analysis}

We measure the nuclear spin Ramsey free evolutions under the electronic spin states $\ket{m_S=0}$, $\ket{m_S=+1}$, and $\ket{m_S=-1}$ in Figs.~\ref{Supp_Fig_Ramsey}(a-c), as well as the double-quantum Ramsey shown in Fig.~\ref{Supp_Fig_Ramsey}(e) and spin echo shown in Fig.~\ref{Supp_Fig_Ramsey}(d). We analyze again the decoherence mechanism and show that by directly comparing the nuclear spin coherence times under different electronic spin states, we can reveal the inhomogeneity and material-dependent ratio $\alpha_Q/\alpha_A$.

Figures.~\ref{Supp_Fig_Ramsey}(b,c) show the nuclear spin Ramsey coherence times for states $(\ket{0}+\ket{\pm1})/\sqrt{2}$ under different electronic spin states $\ket{m_S=\pm 1}$. In particular, the coherence time for nuclear spin state $(\ket{0}+\ket{\pm1})/\sqrt{2}$ under electronic spin state $\ket{m_S=+1}$ is similar to the the coherence time for nuclear spin state $(\ket{0}+\ket{\mp1})/\sqrt{2}$ under electronic spin state $\ket{m_S=-1}$. For both electronic spin states, the nuclear spin coherence times are shorter than the one under the electronic spin state $\ket{m_S=0}$ as shown in Fig.~\ref{Supp_Fig_Ramsey}(a).

We now show that these results are a validation of our model based on the analysis of the noise source. We start by analyzing the double-quantum (DQ) Ramsey measurement shown in Fig.~\ref{Supp_Fig_Ramsey}(e). The nuclear spin  is prepared in the initial state $(\ket{+1}+\ket{-1})/\sqrt{2}$ ($m_S=0$) by simultaneously switching on two RF tones (with the same amplitudes) on resonances with two single-quantum (SQ) nuclear spin transitions $\ket{0}\leftrightarrow\ket{\pm1}$ for a $\pi/\sqrt{2}$ pulse length, immediately after applying a $\pi$ pulse to transfer nuclear spin population from $\ket{+1}$ to $\ket{0}$ (same as Ref.~\cite{jarmola_demonstration_2021}). We note that the state preparation pulse length is directly calibrated by measuring the DQ Rabi oscillation (cycling between $\ket{m_I=0}$ and $(\ket{+1}+\ket{-1})/\sqrt{2}$). As discussed in the main text, for a Lorentzian model, the coherence time is determined by the standard deviation of the transition frequency due to the temperature inhomogeneity in addition to the magnetic field inhomogeneity (assuming non-static sources are negligible). The inhomogeneity in the quadrupole term $\delta Q$ does not affect this superposition and the DQ coherence time can be expressed as
\begin{equation}
    \frac{1}{T_{2n, DQ}^*} \approx \sigma(2\gamma_n B+2m_S \delta A_{zz})=2|\gamma_n|\sigma_B+2|m_S\alpha_A|\sigma_T.
\end{equation}
The SQ Ramsey coherence time can also be analyzed in a similar way such that
\begin{equation}
    \frac{1}{T_{2n, SQ}^*} \approx \sigma(\delta Q+m_Im_S\delta A_{zz}+m_I\gamma_n B)=|(\alpha_Q+m_Im_S\alpha_A)|\sigma_{T}+|\gamma_n|\sigma_B\approx|1+m_Im_S\frac{\alpha_A}{\alpha_Q}|\alpha_Q\sigma_{T}+\frac{1}{2T_{2n,DQ,m_S=0}^*}.
\end{equation}
We note that here $m_I$ is a quantum number representing the used nuclear spin state $(\ket{0}+\ket{m_I})/\sqrt{2}$ with $m_I=\pm1$ and $m_S$ is a quantum number representing the electronic spin state $\ket{m_S}$ with $m_S=0,\pm1$.


As measured in the experiment, the DQ Ramsey coherence time is $T_{2,DQ,m_S=0}^*\approx3.9$~ms.
With the ratio $\alpha_A/\alpha_Q\approx 1/0.18=5.56$ measured in the main text, the ratio between the temperature-induced decoherence rate of nuclear spin states $(\ket{0}+\ket{-1})/\sqrt{2}$ and $(\ket{0}+\ket{+1})/\sqrt{2}$ under the electronic spin states $\ket{m_S=-1}$ is predicted to be $|1+\alpha_A/\alpha_Q|:|1-\alpha_A/\alpha_Q|\approx 1.44$, which is close to the measured ratio $(\frac{1}{107\mu s}-\frac{1}{7.8\text{ms}}):(\frac{1}{157\mu s}-\frac{1}{7.8\text{ms}})\approx 1.48$ (Fig.~\ref{Supp_Fig_Ramsey}(c)). We further compare to Fig.~\ref{Supp_Fig_Ramsey}(a) where the decoherence rate is set by ${1}/{T_{2n}^*}=\alpha_Q\sigma_{T}+1/T_{2,DQ,m_S=0}^*$.
We focus on the nuclear spin coherence time of the initial state $(\ket{0}+\ket{+1})/\sqrt{2}$ and compare the electronic spin state $\ket{m_S=0}$ and $\ket{m_S=+1}$ (Figs.~\ref{Supp_Fig_Ramsey}(a,b)). The measurement results gives a ratio between the temperature-induced decoherence rate under $\ket{m_S=0}$ and $\ket{m_S=+1}$ to be $(\frac{1}{720\mu s}-\frac{1}{7.8\text{ms}}):(\frac{1}{110\mu s}-\frac{1}{7.8\text{ms}})\approx 0.14$, which is consistent with our inhomogeneity analysis predicting $1:|1+\alpha_A/\alpha_Q|\approx0.15$. Thus, by simply measuring the SQ and DQ coherence times and comparing different nuclear and electronic spin states, one can already provide a relatively good estimation of the inhomogeneity of the temperature.

\begin{figure*}[htbp]
\centering \includegraphics[width=\textwidth]{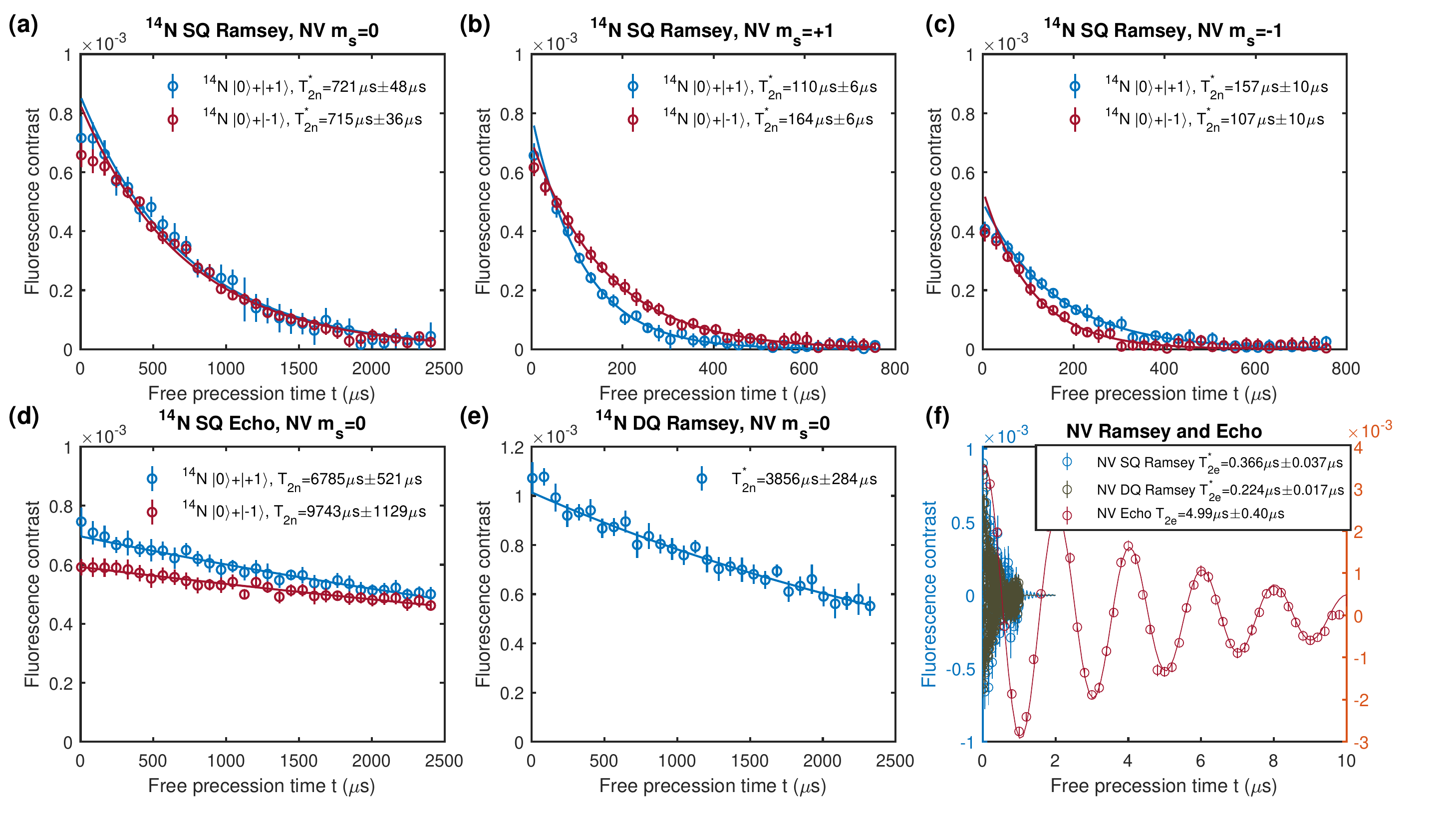}
\caption{\label{Supp_Fig_Ramsey} \textbf{Coherence time measurements.} (a) Nuclear spin Ramsey measurement with NV electronic spin in $\ket{m_S=0}$. The blue and red curves represent the measurement of the initial nuclear spin states $(\ket{m_I=0}+\ket{m_I=+1})/\sqrt{2}$ and  $(\ket{m_I=0}+\ket{m_I=-1})/\sqrt{2}$. The data is fit to the function $S(t)=c_0e^{-t/T_{2n}^*}$ with $c_0$ and $T_{2n}^*$ as the fitting parameters. The first three data points are excluded in the fitting to better capture the long-time exponential behavior. Same fitting rules apply to all the data in (b-e). (b) Nuclear spin Ramsey measurement with NV electronic spin in $\ket{m_S=+1}$. (c) Nuclear spin Ramsey measurement $T_{2n}^*$ with NV electronic spin in $\ket{m_S=-1}$. (d) Nuclear spin Echo measurement with NV electronic spin in $\ket{m_S=0}$. (e) Nuclear spin double-quantum Ramsey measurement. The nuclear spin initial state is prepared to $(\ket{m_I=+1}+\ket{m_I=-1})\sqrt{2}$ with the NV electronic spin in $\ket{m_S=0}$. (f) NV electronic spin Ramsey and Echo measurement. The data is fit to the function $S(t)=c_0e^{-t/T_{2e}}\cos(\omega t)$ with $c_0$, $T_{2e}$ and $\omega$ as the fitting parameters. For both the SQ and DQ Ramsey experiments, we use the nuclear spin level $|m_I=+1\rangle$. While for the Echo experiment, we address both $\ket{m_I=0}$ and $\ket{m_I=+1}$ by applying a short pulse with a large power. For both SQ experiments, we measure the coherence of the electronic spin state $(\ket{m_S=0}+\ket{m_S=-1})/\sqrt{2}$.}
\end{figure*}

Although this work focuses on the study of the nuclear spin coherence time, here we also measure and analyze the electronic spin coherence time and find that the electronic spin SQ Ramsey coherence is also limited by the temperature inhomogeneity. 
The measurement in Ref.~\cite{soshenko_temperature_2020} indicates that $3.6\delta Q/Q=\delta D/D$ for temperature-induced variation, thus we can estimate the temperature contribution in the electronic spin coherence time using
\begin{equation}
    \frac{1}{T_{2e, SQ}^*}\approx\sigma(\delta D)+\gamma_e\sigma_B=\sigma(\delta Q)\frac{\delta D}{\delta Q}+\gamma_e\sigma_B=3.6\frac{D}{Q}\frac{1}{T_{2n,SQ, m_S=0}^*}+\gamma_e\sigma_B\approx\frac{1}{0.345\mu s}+\gamma_e\sigma_B.
\end{equation}
In Fig.~\ref{Supp_Fig_Ramsey}(f), the SQ Ramsey coherence time of the electronic spin is 0.366~$\mu$s and the DQ Ramsey is 0.224$\mu$s.  In principle, one could predict the $\gamma_e \sigma_B$ using the nuclear spin DQ Ramsey coherence time $1/T_{2n, DQ}^*\sim 2\gamma_n\sigma_B$ and ratio $\gamma_e/\gamma_n$, which gives $T_{2e,SQ}^*\sim0.245\mu s$ and $T_{2e,DQ}^*\sim 0.42\mu s$. These predictions match the measurement in order of magnitudes. Thus, the decoherence of the electronic spin is also limited by the temperature inhomogeneity. We note that in NV ensemble samples, the dipolar spin bath due to other spin defects such as P1 centers also contributes to the decoherence of the NV electronic spin~\cite{bauch_decoherence_2020,bauch_ultralong_2018}. Different from the inhomogeneity effects, the spin bath contribution involves the interaction and dynamics within the bath. Further distinguishing and decoupling these noise sources co-existing with the inhomogeneity is outside the scope of this study.  



\begin{figure*}[htbp]
\centering \includegraphics[width=\textwidth]{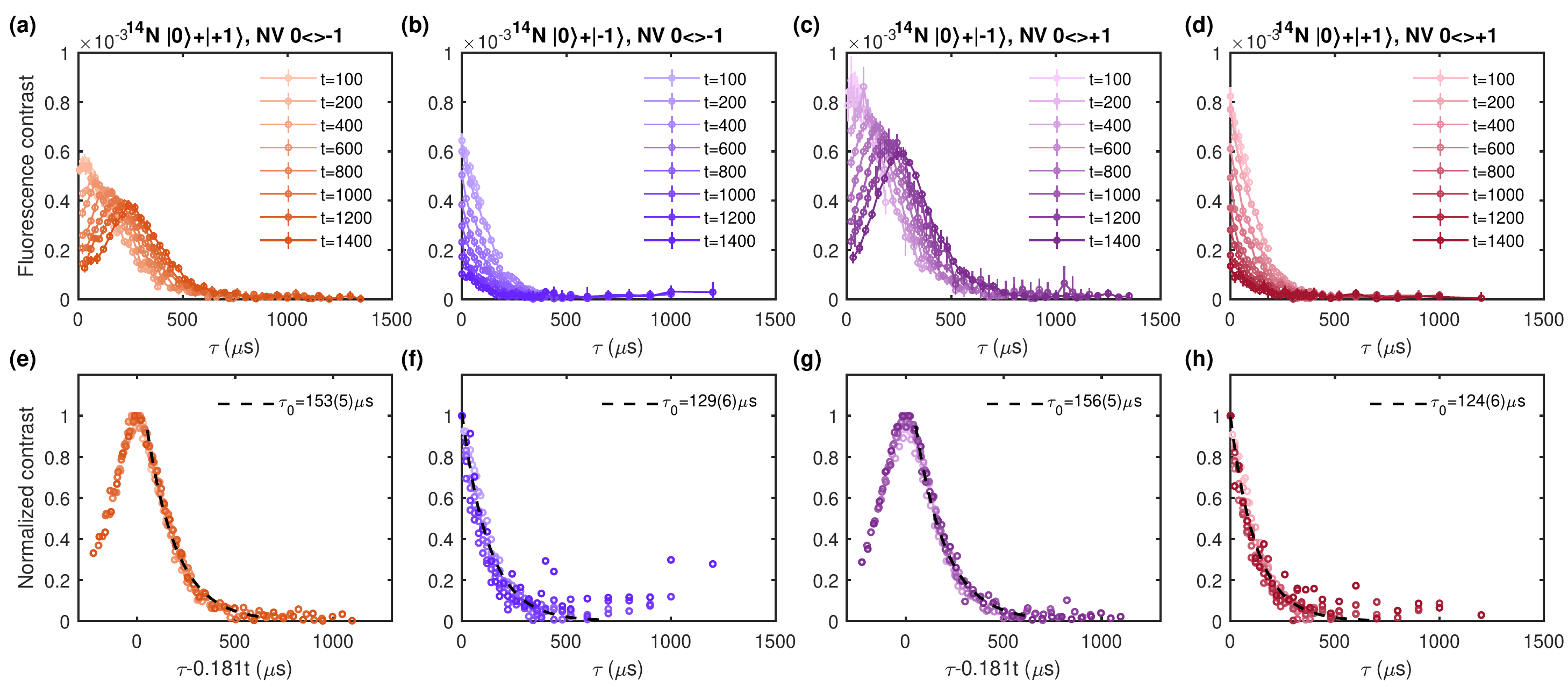}
\caption{\label{Supp_Fig_Echo} \textbf{Pulse location sweep.} (a) Pulse location sweep under different total free precession times $t$. The nuclear spin state is prepared to $(\ket{0}+\ket{+1})/\sqrt{2}$ and the electronic spin flip is between $\ket{m_S=0}$ and $\ket{m_S=-1}$. The fluorescence contrast is measured by sweeping the phase $\phi$ of the final pulse and fitting to a cosine function. (b)  Pulse location sweep under different total free precession times $t$. The nuclear spin state is prepared to $(\ket{0}+\ket{-1})/\sqrt{2}$ and the electronic spin flip is between $\ket{m_S=0}$ and $\ket{m_S=-1}$.  (c)  Pulse location sweep under different total free precession times $t$. The nuclear spin state is prepared to $(\ket{0}+\ket{-1})/\sqrt{2}$ and the electronic spin flip is between $\ket{m_S=0}$ and $\ket{m_S=+1}$. (d)  Pulse location sweep under different total free precession times $t$. The nuclear spin state is prepared to $(\ket{0}+\ket{+1})/\sqrt{2}$ and the electronic spin flip is between $\ket{m_S=0}$ and $\ket{m_S=+1}$. (e-h) Same data as (a-d) but shown with normalization by its maximum values. For (e) and (g), the x axis is also shifted by 0.181t to best show the overlap.}
\end{figure*}

In Fig.~\ref{Supp_Fig_Echo}, we show the complete data with and without normalization for the pulse location sweep experiments for all different combinations of the nuclear spin and electronic spin levels.


In the main text, the ratio predicted in Fig.~2c is a little bit different from the one predicted in Fig.~2d, This is due to the residual decoherence sources that are not considered when calculating the intercept with the $x$ axis in (d). Thus, prediction based on the experiment in (c) is more precise since it directly uses the crossing between the two straight lines to predict the ratio, which does not requires a knowledge of the residual decoherence sources, although such a source can be characterized by other experiments such as DQ Ramsey measurement discussed above.

\section{Temperature-dependence measurement}
All of our experiments are performed at ambient temperature ($\sim300$~K) and pressure. We focus a 0.4~W laser to a small spot with a size around 10-30~$\mu$m. By varying the time duration of the initialization laser pulse, we can vary the temperature and measure its effect on the energy levels of the nuclear and electronic spin.

In Fig.~\ref{Supp_Fig_Resonances}(a), we measure the NV electronic spin transition frequencies and nuclear spin transition frequencies under eight different initialization pulse lengths. We use the electronic spin resonance measurements as a reference to calculate the temperature shift under each sequences with the relation $dD/dT\approx-(2\pi)77.7$~kHz/K~\cite{jarmola_robust_2020}, where the resonance shift is obtained by calculating the average of the two resonance frequencies between $\ket{m_S=0}$ and $\ket{m_S=\pm1}$. Then we keep the same sequence length and experimental repetition to keep the same heat load, and implement the nuclear spin resonance measurements.  In Fig.~\ref{Supp_Fig_Resonances}(b), we show the shifts of the six nuclear spin SQ transition frequencies as a function of temperature. In Fig.~\ref{Supp_Fig_Resonances}(c), we show the calculated quadrupole and hyperfine frequency shifts. The quadrupole shift is obtained by calculating the average of the two transitions between $\ket{m_I=0}$ and $\ket{m_I=\pm1}$ when NV is in a state $\ket{m_S=0}$ such that $\delta Q=\delta\omega_1+\delta\omega_2$. The hyperfine shift is obtained by calculating $\delta |A_{zz}|=\delta\omega_4+\delta\omega_5-\delta\omega_3-\delta\omega_6$. We note that $\omega_1,\cdots,\omega_6$ are the six nuclear spin transitions as shown in Fig.~1a in the main text.

\begin{figure*}[htbp]
\centering \includegraphics[width=\textwidth]{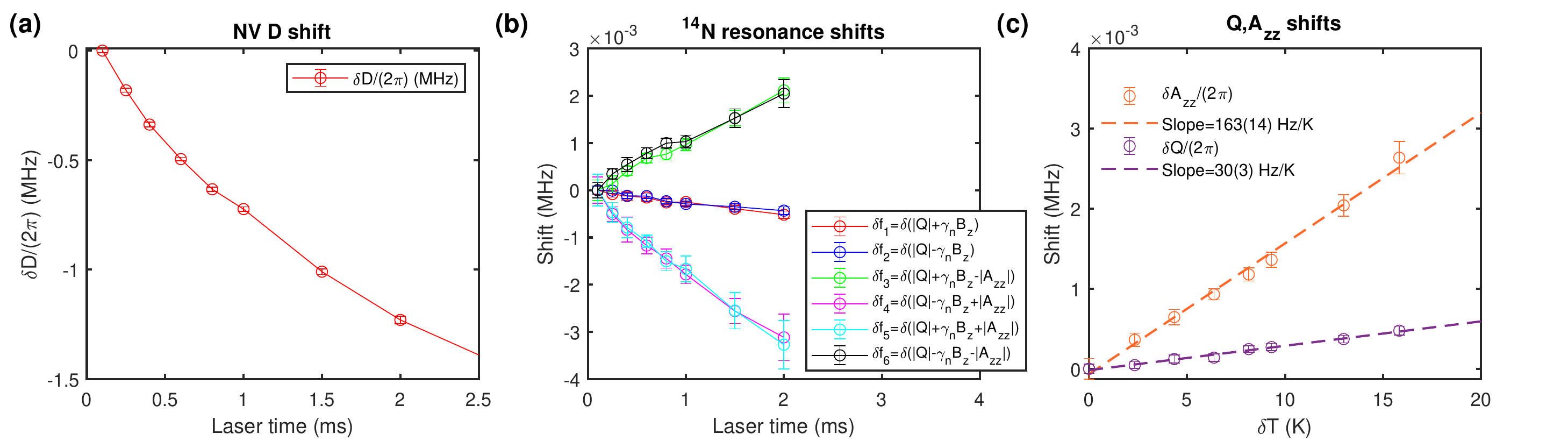}
\caption{\label{Supp_Fig_Resonances} \textbf{Temperature shifts of $D$, $Q$, $A_{zz}$.} (a) Shifts of zero-field splitting under different laser initialization times. (b) Shifts of the six different nuclear spin transitions $f_1,f_2,f_3,f_4,f_5,f_6$. (c) Shifts of the quadrupole and hyperfine interactions calculated with the data in (a) and (b). }
\end{figure*}

\section{Simulation of large temperature fluctuations}
In the main text, we use the first-principles calculation results to numerically study the coherence protection of a large temperature inhomogeneity. In Fig.~\ref{Fig_Simulation}, we further simulate an even larger temperature inhomogeneity to demonstrate the applicability of our method under extreme conditions. For a temperature inhomogeneity satisfying a Lorentzian distribution with a standard deviation $\sigma_{T}=25$~K such that the variation of both quadrupole and hyperfine interactions remains approximately linear, more than $10$~ms coherence time can be achieved (assuming no other noise sources), which shows a 400-fold improvement. For a $\sigma_{T}=100$~K temperature inhomogeneity involving strong nonlinear effects, we still get $\sim3$~ms coherence time showing a 100-fold improvement. 

\begin{figure}[htbp]
\centering \includegraphics[width=0.65\textwidth]{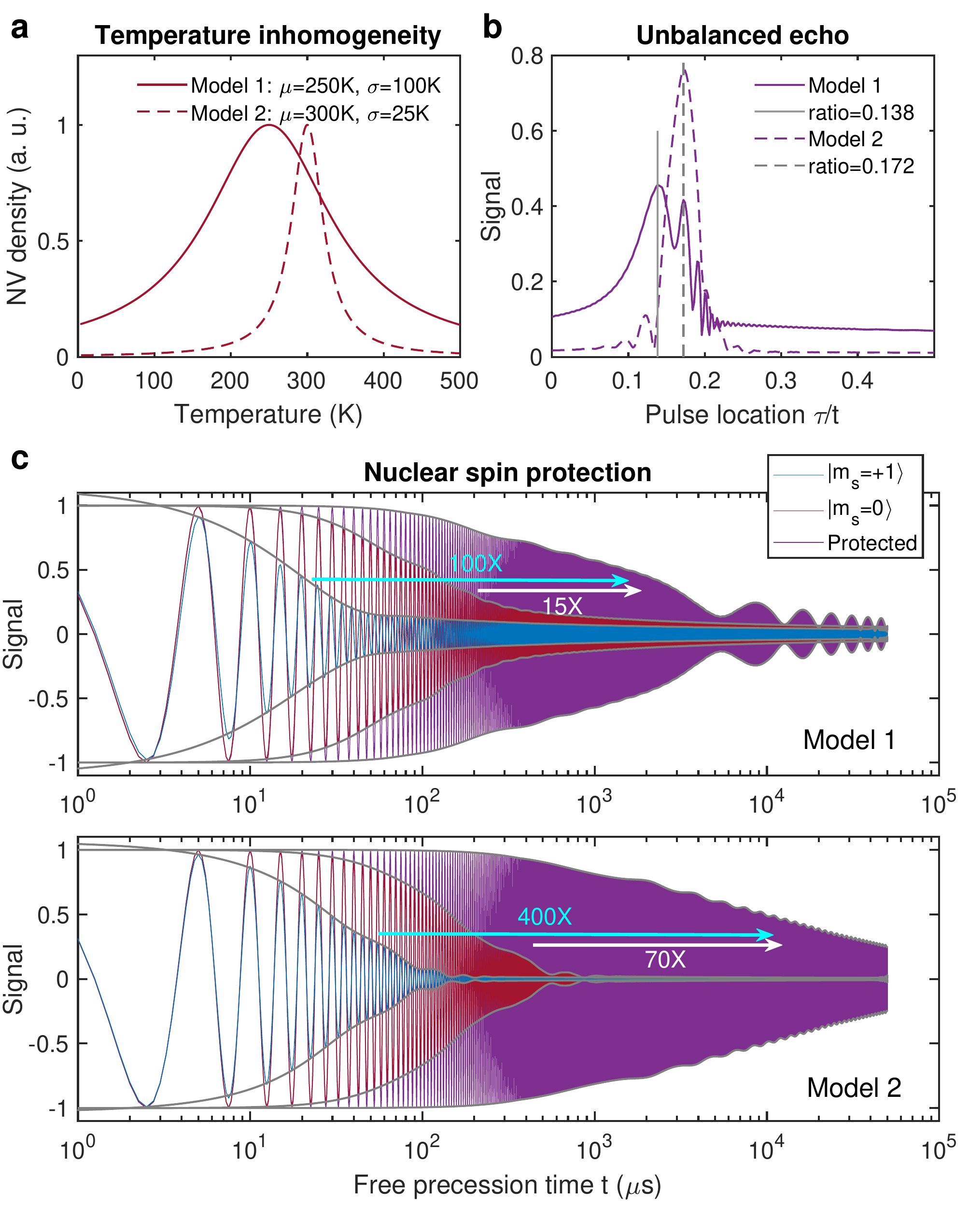}
\caption{\label{Fig_Simulation} \textbf{Simulation of the coherence protection under large temperature inhomogeneities.} a. Temperature distribution in an ensemble of NV center. Both two distributions satisfy a Lorentzian function $f({T})=\frac{1}{\pi\sigma}\frac{\sigma^2}{\sigma^2+({T}-\mu)^2}$. b. Pulse location sweep with $t=2$~ms. The nuclear spin state is chosen to be $(\ket{0}+\ket{-1})/\sqrt{2}$ and the electronic spin flip is between $\ket{m_S=0}$ and $\ket{m_S=+1}$. The optimal ratios are chosen at the maximum phase sweep amplitude. c. Simulated coherence protection for the two models. }
\end{figure}

\end{widetext}

\clearpage
\newpage

\bibliography{main_text}

\begin{thebibliography}{58}%
\makeatletter
\providecommand \@ifxundefined [1]{%
 \@ifx{#1\undefined}
}%
\providecommand \@ifnum [1]{%
 \ifnum #1\expandafter \@firstoftwo
 \else \expandafter \@secondoftwo
 \fi
}%
\providecommand \@ifx [1]{%
 \ifx #1\expandafter \@firstoftwo
 \else \expandafter \@secondoftwo
 \fi
}%
\providecommand \natexlab [1]{#1}%
\providecommand \enquote  [1]{``#1''}%
\providecommand \bibnamefont  [1]{#1}%
\providecommand \bibfnamefont [1]{#1}%
\providecommand \citenamefont [1]{#1}%
\providecommand \href@noop [0]{\@secondoftwo}%
\providecommand \href [0]{\begingroup \@sanitize@url \@href}%
\providecommand \@href[1]{\@@startlink{#1}\@@href}%
\providecommand \@@href[1]{\endgroup#1\@@endlink}%
\providecommand \@sanitize@url [0]{\catcode `\\12\catcode `\$12\catcode
  `\&12\catcode `\#12\catcode `\^12\catcode `\_12\catcode `\%12\relax}%
\providecommand \@@startlink[1]{}%
\providecommand \@@endlink[0]{}%
\providecommand \url  [0]{\begingroup\@sanitize@url \@url }%
\providecommand \@url [1]{\endgroup\@href {#1}{\urlprefix }}%
\providecommand \urlprefix  [0]{URL }%
\providecommand \Eprint [0]{\href }%
\providecommand \doibase [0]{https://doi.org/}%
\providecommand \selectlanguage [0]{\@gobble}%
\providecommand \bibinfo  [0]{\@secondoftwo}%
\providecommand \bibfield  [0]{\@secondoftwo}%
\providecommand \translation [1]{[#1]}%
\providecommand \BibitemOpen [0]{}%
\providecommand \bibitemStop [0]{}%
\providecommand \bibitemNoStop [0]{.\EOS\space}%
\providecommand \EOS [0]{\spacefactor3000\relax}%
\providecommand \BibitemShut  [1]{\csname bibitem#1\endcsname}%
\let\auto@bib@innerbib\@empty
\bibitem [{\citenamefont {Wolfowicz}\ \emph {et~al.}(2021)\citenamefont
  {Wolfowicz}, \citenamefont {Heremans}, \citenamefont {Anderson},
  \citenamefont {Kanai}, \citenamefont {Seo}, \citenamefont {Gali},
  \citenamefont {Galli},\ and\ \citenamefont
  {Awschalom}}]{wolfowicz_quantum_2021}%
  \BibitemOpen
  \bibfield  {author} {\bibinfo {author} {\bibfnamefont {G.}~\bibnamefont
  {Wolfowicz}}, \bibinfo {author} {\bibfnamefont {F.~J.}\ \bibnamefont
  {Heremans}}, \bibinfo {author} {\bibfnamefont {C.~P.}\ \bibnamefont
  {Anderson}}, \bibinfo {author} {\bibfnamefont {S.}~\bibnamefont {Kanai}},
  \bibinfo {author} {\bibfnamefont {H.}~\bibnamefont {Seo}}, \bibinfo {author}
  {\bibfnamefont {A.}~\bibnamefont {Gali}}, \bibinfo {author} {\bibfnamefont
  {G.}~\bibnamefont {Galli}},\ and\ \bibinfo {author} {\bibfnamefont {D.~D.}\
  \bibnamefont {Awschalom}},\ }\bibfield  {title} {\bibinfo {title} {Quantum
  guidelines for solid-state spin defects},\ }\href
  {https://doi.org/10.1038/s41578-021-00306-y} {\bibfield  {journal} {\bibinfo
  {journal} {Nat. Rev. Mater.}\ }\textbf {\bibinfo {volume} {6}},\ \bibinfo
  {pages} {906} (\bibinfo {year} {2021})}\BibitemShut {NoStop}%
\bibitem [{\citenamefont {Ajoy}\ and\ \citenamefont
  {Cappellaro}(2012)}]{ajoy_stable_2012}%
  \BibitemOpen
  \bibfield  {author} {\bibinfo {author} {\bibfnamefont {A.}~\bibnamefont
  {Ajoy}}\ and\ \bibinfo {author} {\bibfnamefont {P.}~\bibnamefont
  {Cappellaro}},\ }\bibfield  {title} {\bibinfo {title} {Stable three-axis
  nuclear-spin gyroscope in diamond},\ }\href
  {https://doi.org/10.1103/PhysRevA.86.062104} {\bibfield  {journal} {\bibinfo
  {journal} {Phys. Rev. A}\ }\textbf {\bibinfo {volume} {86}},\ \bibinfo
  {pages} {062104} (\bibinfo {year} {2012})}\BibitemShut {NoStop}%
\bibitem [{\citenamefont {Jarmola}\ \emph {et~al.}(2021)\citenamefont
  {Jarmola}, \citenamefont {Lourette}, \citenamefont {Acosta}, \citenamefont
  {Birdwell}, \citenamefont {Bl{\"u}mler}, \citenamefont {Budker},
  \citenamefont {Ivanov},\ and\ \citenamefont
  {Malinovsky}}]{jarmola_demonstration_2021}%
  \BibitemOpen
  \bibfield  {author} {\bibinfo {author} {\bibfnamefont {A.}~\bibnamefont
  {Jarmola}}, \bibinfo {author} {\bibfnamefont {S.}~\bibnamefont {Lourette}},
  \bibinfo {author} {\bibfnamefont {V.~M.}\ \bibnamefont {Acosta}}, \bibinfo
  {author} {\bibfnamefont {A.~G.}\ \bibnamefont {Birdwell}}, \bibinfo {author}
  {\bibfnamefont {P.}~\bibnamefont {Bl{\"u}mler}}, \bibinfo {author}
  {\bibfnamefont {D.}~\bibnamefont {Budker}}, \bibinfo {author} {\bibfnamefont
  {T.}~\bibnamefont {Ivanov}},\ and\ \bibinfo {author} {\bibfnamefont {V.~S.}\
  \bibnamefont {Malinovsky}},\ }\bibfield  {title} {\bibinfo {title}
  {Demonstration of diamond nuclear spin gyroscope},\ }\href
  {https://doi.org/10.1126/sciadv.abl3840} {\bibfield  {journal} {\bibinfo
  {journal} {Sci. Adv.}\ }\textbf {\bibinfo {volume} {7}},\ \bibinfo {pages}
  {eabl3840} (\bibinfo {year} {2021})}\BibitemShut {NoStop}%
\bibitem [{\citenamefont {Soshenko}\ \emph {et~al.}(2021)\citenamefont
  {Soshenko}, \citenamefont {Bolshedvorskii}, \citenamefont {Rubinas},
  \citenamefont {Sorokin}, \citenamefont {Smolyaninov}, \citenamefont
  {Vorobyov},\ and\ \citenamefont {Akimov}}]{soshenko_nuclear_2021}%
  \BibitemOpen
  \bibfield  {author} {\bibinfo {author} {\bibfnamefont {V.~V.}\ \bibnamefont
  {Soshenko}}, \bibinfo {author} {\bibfnamefont {S.~V.}\ \bibnamefont
  {Bolshedvorskii}}, \bibinfo {author} {\bibfnamefont {O.}~\bibnamefont
  {Rubinas}}, \bibinfo {author} {\bibfnamefont {V.~N.}\ \bibnamefont
  {Sorokin}}, \bibinfo {author} {\bibfnamefont {A.~N.}\ \bibnamefont
  {Smolyaninov}}, \bibinfo {author} {\bibfnamefont {V.~V.}\ \bibnamefont
  {Vorobyov}},\ and\ \bibinfo {author} {\bibfnamefont {A.~V.}\ \bibnamefont
  {Akimov}},\ }\bibfield  {title} {\bibinfo {title} {Nuclear spin gyroscope
  based on the nitrogen vacancy center in diamond},\ }\href
  {https://doi.org/10.1103/PhysRevLett.126.197702} {\bibfield  {journal}
  {\bibinfo  {journal} {Phys. Rev. Lett.}\ }\textbf {\bibinfo {volume} {126}},\
  \bibinfo {pages} {197702} (\bibinfo {year} {2021})}\BibitemShut {NoStop}%
\bibitem [{\citenamefont {Taylor}\ \emph {et~al.}(2008)\citenamefont {Taylor},
  \citenamefont {Cappellaro}, \citenamefont {Childress}, \citenamefont {Jiang},
  \citenamefont {Budker}, \citenamefont {Hemmer}, \citenamefont {Yacoby},
  \citenamefont {Walsworth},\ and\ \citenamefont
  {Lukin}}]{taylor_high-sensitivity_2008}%
  \BibitemOpen
  \bibfield  {author} {\bibinfo {author} {\bibfnamefont {J.~M.}\ \bibnamefont
  {Taylor}}, \bibinfo {author} {\bibfnamefont {P.}~\bibnamefont {Cappellaro}},
  \bibinfo {author} {\bibfnamefont {L.}~\bibnamefont {Childress}}, \bibinfo
  {author} {\bibfnamefont {L.}~\bibnamefont {Jiang}}, \bibinfo {author}
  {\bibfnamefont {D.}~\bibnamefont {Budker}}, \bibinfo {author} {\bibfnamefont
  {P.~R.}\ \bibnamefont {Hemmer}}, \bibinfo {author} {\bibfnamefont
  {A.}~\bibnamefont {Yacoby}}, \bibinfo {author} {\bibfnamefont
  {R.}~\bibnamefont {Walsworth}},\ and\ \bibinfo {author} {\bibfnamefont
  {M.~D.}\ \bibnamefont {Lukin}},\ }\bibfield  {title} {\bibinfo {title}
  {High-sensitivity diamond magnetometer with nanoscale resolution},\ }\href
  {https://doi.org/10.1038/nphys1075} {\bibfield  {journal} {\bibinfo
  {journal} {Nat. Phys.}\ }\textbf {\bibinfo {volume} {4}},\ \bibinfo {pages}
  {810} (\bibinfo {year} {2008})}\BibitemShut {NoStop}%
\bibitem [{\citenamefont {Maze}\ \emph {et~al.}(2008)\citenamefont {Maze},
  \citenamefont {Stanwix}, \citenamefont {Hodges}, \citenamefont {Hong},
  \citenamefont {Taylor}, \citenamefont {Cappellaro}, \citenamefont {Jiang},
  \citenamefont {Dutt}, \citenamefont {Togan}, \citenamefont {Zibrov},
  \citenamefont {Yacoby}, \citenamefont {Walsworth},\ and\ \citenamefont
  {Lukin}}]{Maze_nature2008}%
  \BibitemOpen
  \bibfield  {author} {\bibinfo {author} {\bibfnamefont {J.~R.}\ \bibnamefont
  {Maze}}, \bibinfo {author} {\bibfnamefont {P.~L.}\ \bibnamefont {Stanwix}},
  \bibinfo {author} {\bibfnamefont {J.~S.}\ \bibnamefont {Hodges}}, \bibinfo
  {author} {\bibfnamefont {S.}~\bibnamefont {Hong}}, \bibinfo {author}
  {\bibfnamefont {J.~M.}\ \bibnamefont {Taylor}}, \bibinfo {author}
  {\bibfnamefont {P.}~\bibnamefont {Cappellaro}}, \bibinfo {author}
  {\bibfnamefont {L.}~\bibnamefont {Jiang}}, \bibinfo {author} {\bibfnamefont
  {M.~V.~G.}\ \bibnamefont {Dutt}}, \bibinfo {author} {\bibfnamefont
  {E.}~\bibnamefont {Togan}}, \bibinfo {author} {\bibfnamefont {A.~S.}\
  \bibnamefont {Zibrov}}, \bibinfo {author} {\bibfnamefont {A.}~\bibnamefont
  {Yacoby}}, \bibinfo {author} {\bibfnamefont {R.~L.}\ \bibnamefont
  {Walsworth}},\ and\ \bibinfo {author} {\bibfnamefont {M.~D.}\ \bibnamefont
  {Lukin}},\ }\bibfield  {title} {\bibinfo {title} {Nanoscale magnetic sensing
  with an individual electronic spin in diamond},\ }\href
  {https://doi.org/10.1038/nature07279} {\bibfield  {journal} {\bibinfo
  {journal} {Nature}\ }\textbf {\bibinfo {volume} {455}},\ \bibinfo {pages}
  {644} (\bibinfo {year} {2008})}\BibitemShut {NoStop}%
\bibitem [{\citenamefont {Dolde}\ \emph {et~al.}(2011)\citenamefont {Dolde},
  \citenamefont {Fedder}, \citenamefont {Doherty}, \citenamefont {N{\"o}bauer},
  \citenamefont {Rempp}, \citenamefont {Balasubramanian}, \citenamefont {Wolf},
  \citenamefont {Reinhard}, \citenamefont {Hollenberg}, \citenamefont
  {Jelezko},\ and\ \citenamefont {Wrachtrup}}]{dolde_electric-field_2011}%
  \BibitemOpen
  \bibfield  {author} {\bibinfo {author} {\bibfnamefont {F.}~\bibnamefont
  {Dolde}}, \bibinfo {author} {\bibfnamefont {H.}~\bibnamefont {Fedder}},
  \bibinfo {author} {\bibfnamefont {M.~W.}\ \bibnamefont {Doherty}}, \bibinfo
  {author} {\bibfnamefont {T.}~\bibnamefont {N{\"o}bauer}}, \bibinfo {author}
  {\bibfnamefont {F.}~\bibnamefont {Rempp}}, \bibinfo {author} {\bibfnamefont
  {G.}~\bibnamefont {Balasubramanian}}, \bibinfo {author} {\bibfnamefont
  {T.}~\bibnamefont {Wolf}}, \bibinfo {author} {\bibfnamefont {F.}~\bibnamefont
  {Reinhard}}, \bibinfo {author} {\bibfnamefont {L.~C.~L.}\ \bibnamefont
  {Hollenberg}}, \bibinfo {author} {\bibfnamefont {F.}~\bibnamefont
  {Jelezko}},\ and\ \bibinfo {author} {\bibfnamefont {J.}~\bibnamefont
  {Wrachtrup}},\ }\bibfield  {title} {\bibinfo {title} {Electric-field sensing
  using single diamond spins},\ }\href {https://doi.org/10.1038/nphys1969}
  {\bibfield  {journal} {\bibinfo  {journal} {Nat. Phys.}\ }\textbf {\bibinfo
  {volume} {7}},\ \bibinfo {pages} {459} (\bibinfo {year} {2011})}\BibitemShut
  {NoStop}%
\bibitem [{\citenamefont {Kucsko}\ \emph {et~al.}(2013)\citenamefont {Kucsko},
  \citenamefont {Maurer}, \citenamefont {Yao}, \citenamefont {Kubo},
  \citenamefont {Noh}, \citenamefont {Lo}, \citenamefont {Park},\ and\
  \citenamefont {Lukin}}]{kucsko_nanometre-scale_2013}%
  \BibitemOpen
  \bibfield  {author} {\bibinfo {author} {\bibfnamefont {G.}~\bibnamefont
  {Kucsko}}, \bibinfo {author} {\bibfnamefont {P.~C.}\ \bibnamefont {Maurer}},
  \bibinfo {author} {\bibfnamefont {N.~Y.}\ \bibnamefont {Yao}}, \bibinfo
  {author} {\bibfnamefont {M.}~\bibnamefont {Kubo}}, \bibinfo {author}
  {\bibfnamefont {H.~J.}\ \bibnamefont {Noh}}, \bibinfo {author} {\bibfnamefont
  {P.~K.}\ \bibnamefont {Lo}}, \bibinfo {author} {\bibfnamefont
  {H.}~\bibnamefont {Park}},\ and\ \bibinfo {author} {\bibfnamefont {M.~D.}\
  \bibnamefont {Lukin}},\ }\bibfield  {title} {\bibinfo {title}
  {Nanometre-scale thermometry in a living cell},\ }\href
  {https://doi.org/10.1038/nature12373} {\bibfield  {journal} {\bibinfo
  {journal} {Nature}\ }\textbf {\bibinfo {volume} {500}},\ \bibinfo {pages}
  {54} (\bibinfo {year} {2013})}\BibitemShut {NoStop}%
\bibitem [{\citenamefont {Choi}\ \emph {et~al.}(2020)\citenamefont {Choi},
  \citenamefont {Zhou}, \citenamefont {Landig}, \citenamefont {Wu},
  \citenamefont {Yu}, \citenamefont {Stetina}, \citenamefont {Kucsko},
  \citenamefont {Mango}, \citenamefont {Needleman}, \citenamefont {Samuel},
  \citenamefont {Maurer}, \citenamefont {Park},\ and\ \citenamefont
  {Lukin}}]{Choi2020}%
  \BibitemOpen
  \bibfield  {author} {\bibinfo {author} {\bibfnamefont {J.}~\bibnamefont
  {Choi}}, \bibinfo {author} {\bibfnamefont {H.}~\bibnamefont {Zhou}}, \bibinfo
  {author} {\bibfnamefont {R.}~\bibnamefont {Landig}}, \bibinfo {author}
  {\bibfnamefont {H.-Y.}\ \bibnamefont {Wu}}, \bibinfo {author} {\bibfnamefont
  {X.}~\bibnamefont {Yu}}, \bibinfo {author} {\bibfnamefont {S.~E.~V.}\
  \bibnamefont {Stetina}}, \bibinfo {author} {\bibfnamefont {G.}~\bibnamefont
  {Kucsko}}, \bibinfo {author} {\bibfnamefont {S.~E.}\ \bibnamefont {Mango}},
  \bibinfo {author} {\bibfnamefont {D.~J.}\ \bibnamefont {Needleman}}, \bibinfo
  {author} {\bibfnamefont {A.~D.~T.}\ \bibnamefont {Samuel}}, \bibinfo {author}
  {\bibfnamefont {P.~C.}\ \bibnamefont {Maurer}}, \bibinfo {author}
  {\bibfnamefont {H.}~\bibnamefont {Park}},\ and\ \bibinfo {author}
  {\bibfnamefont {M.~D.}\ \bibnamefont {Lukin}},\ }\bibfield  {title} {\bibinfo
  {title} {Probing and manipulating embryogenesis via nanoscale thermometry and
  temperature control},\ }\href {https://doi.org/10.1073/pnas.1922730117}
  {\bibfield  {journal} {\bibinfo  {journal} {Proc. Natl. Acad. Sci. U.S.A.}\
  }\textbf {\bibinfo {volume} {117}},\ \bibinfo {pages} {14636} (\bibinfo
  {year} {2020})}\BibitemShut {NoStop}%
\bibitem [{\citenamefont {Doherty}\ \emph
  {et~al.}(2014{\natexlab{a}})\citenamefont {Doherty}, \citenamefont
  {Struzhkin}, \citenamefont {Simpson}, \citenamefont {McGuinness},
  \citenamefont {Meng}, \citenamefont {Stacey}, \citenamefont {Karle},
  \citenamefont {Hemley}, \citenamefont {Manson}, \citenamefont {Hollenberg},\
  and\ \citenamefont {Prawer}}]{doherty_electronic_2014}%
  \BibitemOpen
  \bibfield  {author} {\bibinfo {author} {\bibfnamefont {M.~W.}\ \bibnamefont
  {Doherty}}, \bibinfo {author} {\bibfnamefont {V.~V.}\ \bibnamefont
  {Struzhkin}}, \bibinfo {author} {\bibfnamefont {D.~A.}\ \bibnamefont
  {Simpson}}, \bibinfo {author} {\bibfnamefont {L.~P.}\ \bibnamefont
  {McGuinness}}, \bibinfo {author} {\bibfnamefont {Y.}~\bibnamefont {Meng}},
  \bibinfo {author} {\bibfnamefont {A.}~\bibnamefont {Stacey}}, \bibinfo
  {author} {\bibfnamefont {T.~J.}\ \bibnamefont {Karle}}, \bibinfo {author}
  {\bibfnamefont {R.~J.}\ \bibnamefont {Hemley}}, \bibinfo {author}
  {\bibfnamefont {N.~B.}\ \bibnamefont {Manson}}, \bibinfo {author}
  {\bibfnamefont {L.~C.~L.}\ \bibnamefont {Hollenberg}},\ and\ \bibinfo
  {author} {\bibfnamefont {S.}~\bibnamefont {Prawer}},\ }\bibfield  {title}
  {\bibinfo {title} {Electronic {{Properties}} and {{Metrology Applications}}
  of the {{Diamond NV}} - {{Center}} under {{Pressure}}},\ }\href
  {https://doi.org/10.1103/PhysRevLett.112.047601} {\bibfield  {journal}
  {\bibinfo  {journal} {Phys. Rev. Lett.}\ }\textbf {\bibinfo {volume} {112}},\
  \bibinfo {pages} {047601} (\bibinfo {year} {2014}{\natexlab{a}})}\BibitemShut
  {NoStop}%
\bibitem [{\citenamefont {Yip}\ \emph {et~al.}(2019)\citenamefont {Yip},
  \citenamefont {Ho}, \citenamefont {Yu}, \citenamefont {Chen}, \citenamefont
  {Zhang}, \citenamefont {Kasahara}, \citenamefont {Mizukami}, \citenamefont
  {Shibauchi}, \citenamefont {Matsuda}, \citenamefont {Goh},\ and\
  \citenamefont {Yang}}]{Yip2019}%
  \BibitemOpen
  \bibfield  {author} {\bibinfo {author} {\bibfnamefont {K.~Y.}\ \bibnamefont
  {Yip}}, \bibinfo {author} {\bibfnamefont {K.~O.}\ \bibnamefont {Ho}},
  \bibinfo {author} {\bibfnamefont {K.~Y.}\ \bibnamefont {Yu}}, \bibinfo
  {author} {\bibfnamefont {Y.}~\bibnamefont {Chen}}, \bibinfo {author}
  {\bibfnamefont {W.}~\bibnamefont {Zhang}}, \bibinfo {author} {\bibfnamefont
  {S.}~\bibnamefont {Kasahara}}, \bibinfo {author} {\bibfnamefont
  {Y.}~\bibnamefont {Mizukami}}, \bibinfo {author} {\bibfnamefont
  {T.}~\bibnamefont {Shibauchi}}, \bibinfo {author} {\bibfnamefont
  {Y.}~\bibnamefont {Matsuda}}, \bibinfo {author} {\bibfnamefont {S.~K.}\
  \bibnamefont {Goh}},\ and\ \bibinfo {author} {\bibfnamefont {S.}~\bibnamefont
  {Yang}},\ }\bibfield  {title} {\bibinfo {title} {Measuring magnetic field
  texture in correlated electron systems under extreme conditions},\ }\href
  {https://doi.org/10.1126/science.aaw4278} {\bibfield  {journal} {\bibinfo
  {journal} {Science}\ }\textbf {\bibinfo {volume} {366}},\ \bibinfo {pages}
  {1355} (\bibinfo {year} {2019})}\BibitemShut {NoStop}%
\bibitem [{\citenamefont {Lesik}\ \emph {et~al.}(2019)\citenamefont {Lesik},
  \citenamefont {Plisson}, \citenamefont {Toraille}, \citenamefont {Renaud},
  \citenamefont {Occelli}, \citenamefont {Schmidt}, \citenamefont {Salord},
  \citenamefont {Delobbe}, \citenamefont {Debuisschert}, \citenamefont
  {Rondin}, \citenamefont {Loubeyre},\ and\ \citenamefont {Roch}}]{Lesik2019}%
  \BibitemOpen
  \bibfield  {author} {\bibinfo {author} {\bibfnamefont {M.}~\bibnamefont
  {Lesik}}, \bibinfo {author} {\bibfnamefont {T.}~\bibnamefont {Plisson}},
  \bibinfo {author} {\bibfnamefont {L.}~\bibnamefont {Toraille}}, \bibinfo
  {author} {\bibfnamefont {J.}~\bibnamefont {Renaud}}, \bibinfo {author}
  {\bibfnamefont {F.}~\bibnamefont {Occelli}}, \bibinfo {author} {\bibfnamefont
  {M.}~\bibnamefont {Schmidt}}, \bibinfo {author} {\bibfnamefont
  {O.}~\bibnamefont {Salord}}, \bibinfo {author} {\bibfnamefont
  {A.}~\bibnamefont {Delobbe}}, \bibinfo {author} {\bibfnamefont
  {T.}~\bibnamefont {Debuisschert}}, \bibinfo {author} {\bibfnamefont
  {L.}~\bibnamefont {Rondin}}, \bibinfo {author} {\bibfnamefont
  {P.}~\bibnamefont {Loubeyre}},\ and\ \bibinfo {author} {\bibfnamefont
  {J.-F.}\ \bibnamefont {Roch}},\ }\bibfield  {title} {\bibinfo {title}
  {Magnetic measurements on micrometer-sized samples under high pressure using
  designed nv centers},\ }\href {https://doi.org/10.1126/science.aaw4329}
  {\bibfield  {journal} {\bibinfo  {journal} {Science}\ }\textbf {\bibinfo
  {volume} {366}},\ \bibinfo {pages} {1359} (\bibinfo {year}
  {2019})}\BibitemShut {NoStop}%
\bibitem [{\citenamefont {Hsieh}\ \emph {et~al.}(2019)\citenamefont {Hsieh},
  \citenamefont {Bhattacharyya}, \citenamefont {Zu}, \citenamefont {Mittiga},
  \citenamefont {Smart}, \citenamefont {Machado}, \citenamefont {Kobrin},
  \citenamefont {H{\"o}hn}, \citenamefont {Rui}, \citenamefont {Kamrani},
  \citenamefont {Chatterjee}, \citenamefont {Choi}, \citenamefont {Zaletel},
  \citenamefont {Struzhkin}, \citenamefont {Moore}, \citenamefont {Levitas},
  \citenamefont {Jeanloz},\ and\ \citenamefont {Yao}}]{Hsieh2019}%
  \BibitemOpen
  \bibfield  {author} {\bibinfo {author} {\bibfnamefont {S.}~\bibnamefont
  {Hsieh}}, \bibinfo {author} {\bibfnamefont {P.}~\bibnamefont
  {Bhattacharyya}}, \bibinfo {author} {\bibfnamefont {C.}~\bibnamefont {Zu}},
  \bibinfo {author} {\bibfnamefont {T.}~\bibnamefont {Mittiga}}, \bibinfo
  {author} {\bibfnamefont {T.~J.}\ \bibnamefont {Smart}}, \bibinfo {author}
  {\bibfnamefont {F.}~\bibnamefont {Machado}}, \bibinfo {author} {\bibfnamefont
  {B.}~\bibnamefont {Kobrin}}, \bibinfo {author} {\bibfnamefont {T.~O.}\
  \bibnamefont {H{\"o}hn}}, \bibinfo {author} {\bibfnamefont {N.~Z.}\
  \bibnamefont {Rui}}, \bibinfo {author} {\bibfnamefont {M.}~\bibnamefont
  {Kamrani}}, \bibinfo {author} {\bibfnamefont {S.}~\bibnamefont {Chatterjee}},
  \bibinfo {author} {\bibfnamefont {S.}~\bibnamefont {Choi}}, \bibinfo {author}
  {\bibfnamefont {M.}~\bibnamefont {Zaletel}}, \bibinfo {author} {\bibfnamefont
  {V.~V.}\ \bibnamefont {Struzhkin}}, \bibinfo {author} {\bibfnamefont {J.~E.}\
  \bibnamefont {Moore}}, \bibinfo {author} {\bibfnamefont {V.~I.}\ \bibnamefont
  {Levitas}}, \bibinfo {author} {\bibfnamefont {R.}~\bibnamefont {Jeanloz}},\
  and\ \bibinfo {author} {\bibfnamefont {N.~Y.}\ \bibnamefont {Yao}},\
  }\bibfield  {title} {\bibinfo {title} {Imaging stress and magnetism at high
  pressures using a nanoscale quantum sensor},\ }\href
  {https://doi.org/10.1126/science.aaw4352} {\bibfield  {journal} {\bibinfo
  {journal} {Science}\ }\textbf {\bibinfo {volume} {366}},\ \bibinfo {pages}
  {1349} (\bibinfo {year} {2019})}\BibitemShut {NoStop}%
\bibitem [{\citenamefont {Degen}\ \emph {et~al.}(2017)\citenamefont {Degen},
  \citenamefont {Reinhard},\ and\ \citenamefont
  {Cappellaro}}]{degen_quantum_2017}%
  \BibitemOpen
  \bibfield  {author} {\bibinfo {author} {\bibfnamefont {C.~L.}\ \bibnamefont
  {Degen}}, \bibinfo {author} {\bibfnamefont {F.}~\bibnamefont {Reinhard}},\
  and\ \bibinfo {author} {\bibfnamefont {P.}~\bibnamefont {Cappellaro}},\
  }\bibfield  {title} {\bibinfo {title} {Quantum sensing},\ }\href
  {https://doi.org/10.1103/RevModPhys.89.035002} {\bibfield  {journal}
  {\bibinfo  {journal} {Rev. Mod. Phys.}\ }\textbf {\bibinfo {volume} {89}},\
  \bibinfo {pages} {035002} (\bibinfo {year} {2017})}\BibitemShut {NoStop}%
\bibitem [{\citenamefont {Morton}\ \emph {et~al.}(2008)\citenamefont {Morton},
  \citenamefont {Tyryshkin}, \citenamefont {Brown}, \citenamefont {Shankar},
  \citenamefont {Lovett}, \citenamefont {Ardavan}, \citenamefont {Schenkel},
  \citenamefont {Haller}, \citenamefont {Ager},\ and\ \citenamefont
  {Lyon}}]{morton_solid-state_2008}%
  \BibitemOpen
  \bibfield  {author} {\bibinfo {author} {\bibfnamefont {J.~J.~L.}\
  \bibnamefont {Morton}}, \bibinfo {author} {\bibfnamefont {A.~M.}\
  \bibnamefont {Tyryshkin}}, \bibinfo {author} {\bibfnamefont {R.~M.}\
  \bibnamefont {Brown}}, \bibinfo {author} {\bibfnamefont {S.}~\bibnamefont
  {Shankar}}, \bibinfo {author} {\bibfnamefont {B.~W.}\ \bibnamefont {Lovett}},
  \bibinfo {author} {\bibfnamefont {A.}~\bibnamefont {Ardavan}}, \bibinfo
  {author} {\bibfnamefont {T.}~\bibnamefont {Schenkel}}, \bibinfo {author}
  {\bibfnamefont {E.~E.}\ \bibnamefont {Haller}}, \bibinfo {author}
  {\bibfnamefont {J.~W.}\ \bibnamefont {Ager}},\ and\ \bibinfo {author}
  {\bibfnamefont {S.~A.}\ \bibnamefont {Lyon}},\ }\bibfield  {title} {\bibinfo
  {title} {Solid-state quantum memory using the {{31P}} nuclear spin},\ }\href
  {https://doi.org/10.1038/nature07295} {\bibfield  {journal} {\bibinfo
  {journal} {Nature}\ }\textbf {\bibinfo {volume} {455}},\ \bibinfo {pages}
  {1085} (\bibinfo {year} {2008})}\BibitemShut {NoStop}%
\bibitem [{\citenamefont {Bourassa}\ \emph {et~al.}(2020)\citenamefont
  {Bourassa}, \citenamefont {Anderson}, \citenamefont {Miao}, \citenamefont
  {Onizhuk}, \citenamefont {Ma}, \citenamefont {Crook}, \citenamefont {Abe},
  \citenamefont {{Ul-Hassan}}, \citenamefont {Ohshima}, \citenamefont {Son},
  \citenamefont {Galli},\ and\ \citenamefont
  {Awschalom}}]{bourassa_entanglement_2020}%
  \BibitemOpen
  \bibfield  {author} {\bibinfo {author} {\bibfnamefont {A.}~\bibnamefont
  {Bourassa}}, \bibinfo {author} {\bibfnamefont {C.~P.}\ \bibnamefont
  {Anderson}}, \bibinfo {author} {\bibfnamefont {K.~C.}\ \bibnamefont {Miao}},
  \bibinfo {author} {\bibfnamefont {M.}~\bibnamefont {Onizhuk}}, \bibinfo
  {author} {\bibfnamefont {H.}~\bibnamefont {Ma}}, \bibinfo {author}
  {\bibfnamefont {A.~L.}\ \bibnamefont {Crook}}, \bibinfo {author}
  {\bibfnamefont {H.}~\bibnamefont {Abe}}, \bibinfo {author} {\bibfnamefont
  {J.}~\bibnamefont {{Ul-Hassan}}}, \bibinfo {author} {\bibfnamefont
  {T.}~\bibnamefont {Ohshima}}, \bibinfo {author} {\bibfnamefont {N.~T.}\
  \bibnamefont {Son}}, \bibinfo {author} {\bibfnamefont {G.}~\bibnamefont
  {Galli}},\ and\ \bibinfo {author} {\bibfnamefont {D.~D.}\ \bibnamefont
  {Awschalom}},\ }\bibfield  {title} {\bibinfo {title} {Entanglement and
  control of single nuclear spins in isotopically engineered silicon carbide},\
  }\href {https://doi.org/10.1038/s41563-020-00802-6} {\bibfield  {journal}
  {\bibinfo  {journal} {Nat. Mater.}\ }\textbf {\bibinfo {volume} {19}},\
  \bibinfo {pages} {1319} (\bibinfo {year} {2020})}\BibitemShut {NoStop}%
\bibitem [{\citenamefont {Zwanenburg}\ \emph {et~al.}(2013)\citenamefont
  {Zwanenburg}, \citenamefont {Dzurak}, \citenamefont {Morello}, \citenamefont
  {Simmons}, \citenamefont {Hollenberg}, \citenamefont {Klimeck}, \citenamefont
  {Rogge}, \citenamefont {Coppersmith},\ and\ \citenamefont
  {Eriksson}}]{zwanenburg_silicon_2013}%
  \BibitemOpen
  \bibfield  {author} {\bibinfo {author} {\bibfnamefont {F.~A.}\ \bibnamefont
  {Zwanenburg}}, \bibinfo {author} {\bibfnamefont {A.~S.}\ \bibnamefont
  {Dzurak}}, \bibinfo {author} {\bibfnamefont {A.}~\bibnamefont {Morello}},
  \bibinfo {author} {\bibfnamefont {M.~Y.}\ \bibnamefont {Simmons}}, \bibinfo
  {author} {\bibfnamefont {L.~C.~L.}\ \bibnamefont {Hollenberg}}, \bibinfo
  {author} {\bibfnamefont {G.}~\bibnamefont {Klimeck}}, \bibinfo {author}
  {\bibfnamefont {S.}~\bibnamefont {Rogge}}, \bibinfo {author} {\bibfnamefont
  {S.~N.}\ \bibnamefont {Coppersmith}},\ and\ \bibinfo {author} {\bibfnamefont
  {M.~A.}\ \bibnamefont {Eriksson}},\ }\bibfield  {title} {\bibinfo {title}
  {Silicon quantum electronics},\ }\href
  {https://doi.org/10.1103/RevModPhys.85.961} {\bibfield  {journal} {\bibinfo
  {journal} {Rev. Mod. Phys.}\ }\textbf {\bibinfo {volume} {85}},\ \bibinfo
  {pages} {961} (\bibinfo {year} {2013})}\BibitemShut {NoStop}%
\bibitem [{\citenamefont {Morello}\ \emph {et~al.}(2020)\citenamefont
  {Morello}, \citenamefont {Pla}, \citenamefont {Bertet},\ and\ \citenamefont
  {Jamieson}}]{morello_donor_2020}%
  \BibitemOpen
  \bibfield  {author} {\bibinfo {author} {\bibfnamefont {A.}~\bibnamefont
  {Morello}}, \bibinfo {author} {\bibfnamefont {J.~J.}\ \bibnamefont {Pla}},
  \bibinfo {author} {\bibfnamefont {P.}~\bibnamefont {Bertet}},\ and\ \bibinfo
  {author} {\bibfnamefont {D.~N.}\ \bibnamefont {Jamieson}},\ }\bibfield
  {title} {\bibinfo {title} {Donor {{Spins}} in {{Silicon}} for {{Quantum
  Technologies}}},\ }\href {https://doi.org/10.1002/qute.202000005} {\bibfield
  {journal} {\bibinfo  {journal} {Adv. Quantum Technol.}\ }\textbf {\bibinfo
  {volume} {3}},\ \bibinfo {pages} {2000005} (\bibinfo {year}
  {2020})}\BibitemShut {NoStop}%
\bibitem [{\citenamefont {Ruskuc}\ \emph {et~al.}(2022)\citenamefont {Ruskuc},
  \citenamefont {Wu}, \citenamefont {Rochman}, \citenamefont {Choi},\ and\
  \citenamefont {Faraon}}]{ruskuc_nuclear_2022}%
  \BibitemOpen
  \bibfield  {author} {\bibinfo {author} {\bibfnamefont {A.}~\bibnamefont
  {Ruskuc}}, \bibinfo {author} {\bibfnamefont {C.-J.}\ \bibnamefont {Wu}},
  \bibinfo {author} {\bibfnamefont {J.}~\bibnamefont {Rochman}}, \bibinfo
  {author} {\bibfnamefont {J.}~\bibnamefont {Choi}},\ and\ \bibinfo {author}
  {\bibfnamefont {A.}~\bibnamefont {Faraon}},\ }\bibfield  {title} {\bibinfo
  {title} {Nuclear spin-wave quantum register for a solid-state qubit},\ }\href
  {https://doi.org/10.1038/s41586-021-04293-6} {\bibfield  {journal} {\bibinfo
  {journal} {Nature}\ }\textbf {\bibinfo {volume} {602}},\ \bibinfo {pages}
  {408} (\bibinfo {year} {2022})}\BibitemShut {NoStop}%
\bibitem [{\citenamefont {Suter}\ and\ \citenamefont
  {{\'A}lvarez}(2016)}]{suter_colloquium_2016}%
  \BibitemOpen
  \bibfield  {author} {\bibinfo {author} {\bibfnamefont {D.}~\bibnamefont
  {Suter}}\ and\ \bibinfo {author} {\bibfnamefont {G.~A.}\ \bibnamefont
  {{\'A}lvarez}},\ }\bibfield  {title} {\bibinfo {title} {{\emph{Colloquium}} :
  Protecting quantum information against environmental noise},\ }\href
  {https://doi.org/10.1103/RevModPhys.88.041001} {\bibfield  {journal}
  {\bibinfo  {journal} {Rev. Mod. Phys.}\ }\textbf {\bibinfo {volume} {88}},\
  \bibinfo {pages} {041001} (\bibinfo {year} {2016})}\BibitemShut {NoStop}%
\bibitem [{\citenamefont {Hahn}(1950)}]{hahn_spin_1950}%
  \BibitemOpen
  \bibfield  {author} {\bibinfo {author} {\bibfnamefont {E.~L.}\ \bibnamefont
  {Hahn}},\ }\bibfield  {title} {\bibinfo {title} {Spin {{Echoes}}},\ }\href
  {https://doi.org/10.1103/PhysRev.80.580} {\bibfield  {journal} {\bibinfo
  {journal} {Phys. Rev.}\ }\textbf {\bibinfo {volume} {80}},\ \bibinfo {pages}
  {580} (\bibinfo {year} {1950})}\BibitemShut {NoStop}%
\bibitem [{\citenamefont {Bauch}\ \emph {et~al.}(2018)\citenamefont {Bauch},
  \citenamefont {Hart}, \citenamefont {Schloss}, \citenamefont {Turner},
  \citenamefont {Barry}, \citenamefont {Kehayias}, \citenamefont {Singh},\ and\
  \citenamefont {Walsworth}}]{bauch_ultralong_2018}%
  \BibitemOpen
  \bibfield  {author} {\bibinfo {author} {\bibfnamefont {E.}~\bibnamefont
  {Bauch}}, \bibinfo {author} {\bibfnamefont {C.~A.}\ \bibnamefont {Hart}},
  \bibinfo {author} {\bibfnamefont {J.~M.}\ \bibnamefont {Schloss}}, \bibinfo
  {author} {\bibfnamefont {M.~J.}\ \bibnamefont {Turner}}, \bibinfo {author}
  {\bibfnamefont {J.~F.}\ \bibnamefont {Barry}}, \bibinfo {author}
  {\bibfnamefont {P.}~\bibnamefont {Kehayias}}, \bibinfo {author}
  {\bibfnamefont {S.}~\bibnamefont {Singh}},\ and\ \bibinfo {author}
  {\bibfnamefont {R.~L.}\ \bibnamefont {Walsworth}},\ }\bibfield  {title}
  {\bibinfo {title} {Ultralong {{Dephasing Times}} in {{Solid-State Spin
  Ensembles}} via {{Quantum Control}}},\ }\href
  {https://doi.org/10.1103/PhysRevX.8.031025} {\bibfield  {journal} {\bibinfo
  {journal} {Phys. Rev. X}\ }\textbf {\bibinfo {volume} {8}},\ \bibinfo {pages}
  {031025} (\bibinfo {year} {2018})}\BibitemShut {NoStop}%
\bibitem [{\citenamefont {Knowles}\ \emph {et~al.}(2014)\citenamefont
  {Knowles}, \citenamefont {Kara},\ and\ \citenamefont
  {Atat{\"u}re}}]{knowles_observing_2014}%
  \BibitemOpen
  \bibfield  {author} {\bibinfo {author} {\bibfnamefont {H.~S.}\ \bibnamefont
  {Knowles}}, \bibinfo {author} {\bibfnamefont {D.~M.}\ \bibnamefont {Kara}},\
  and\ \bibinfo {author} {\bibfnamefont {M.}~\bibnamefont {Atat{\"u}re}},\
  }\bibfield  {title} {\bibinfo {title} {Observing bulk diamond spin coherence
  in high-purity nanodiamonds},\ }\href {https://doi.org/10.1038/nmat3805}
  {\bibfield  {journal} {\bibinfo  {journal} {Nat. Mater.}\ }\textbf {\bibinfo
  {volume} {13}},\ \bibinfo {pages} {21} (\bibinfo {year} {2014})}\BibitemShut
  {NoStop}%
\bibitem [{\citenamefont {Chen}\ \emph {et~al.}(2018)\citenamefont {Chen},
  \citenamefont {Sun}, \citenamefont {Saha}, \citenamefont {Jaskula},\ and\
  \citenamefont {Cappellaro}}]{chen_protecting_2018}%
  \BibitemOpen
  \bibfield  {author} {\bibinfo {author} {\bibfnamefont {M.}~\bibnamefont
  {Chen}}, \bibinfo {author} {\bibfnamefont {W.~K.~C.}\ \bibnamefont {Sun}},
  \bibinfo {author} {\bibfnamefont {K.}~\bibnamefont {Saha}}, \bibinfo {author}
  {\bibfnamefont {J.-C.}\ \bibnamefont {Jaskula}},\ and\ \bibinfo {author}
  {\bibfnamefont {P.}~\bibnamefont {Cappellaro}},\ }\bibfield  {title}
  {\bibinfo {title} {Protecting solid-state spins from a strongly coupled
  environment},\ }\href {https://doi.org/10.1088/1367-2630/aac542} {\bibfield
  {journal} {\bibinfo  {journal} {New J. Phys.}\ }\textbf {\bibinfo {volume}
  {20}},\ \bibinfo {pages} {063011} (\bibinfo {year} {2018})}\BibitemShut
  {NoStop}%
\bibitem [{\citenamefont {Hart}\ \emph {et~al.}(2021)\citenamefont {Hart},
  \citenamefont {Schloss}, \citenamefont {Turner}, \citenamefont {Scheidegger},
  \citenamefont {Bauch},\ and\ \citenamefont {Walsworth}}]{hart_n_2021}%
  \BibitemOpen
  \bibfield  {author} {\bibinfo {author} {\bibfnamefont {C.~A.}\ \bibnamefont
  {Hart}}, \bibinfo {author} {\bibfnamefont {J.~M.}\ \bibnamefont {Schloss}},
  \bibinfo {author} {\bibfnamefont {M.~J.}\ \bibnamefont {Turner}}, \bibinfo
  {author} {\bibfnamefont {P.~J.}\ \bibnamefont {Scheidegger}}, \bibinfo
  {author} {\bibfnamefont {E.}~\bibnamefont {Bauch}},\ and\ \bibinfo {author}
  {\bibfnamefont {R.~L.}\ \bibnamefont {Walsworth}},\ }\bibfield  {title}
  {\bibinfo {title} {N - {{V}} \textendash{{Diamond Magnetic Microscopy Using}}
  a {{Double Quantum}} 4-{{Ramsey Protocol}}},\ }\href
  {https://doi.org/10.1103/PhysRevApplied.15.044020} {\bibfield  {journal}
  {\bibinfo  {journal} {Phys. Rev. Appl.}\ }\textbf {\bibinfo {volume} {15}},\
  \bibinfo {pages} {044020} (\bibinfo {year} {2021})}\BibitemShut {NoStop}%
\bibitem [{\citenamefont {Reiserer}\ \emph {et~al.}(2016)\citenamefont
  {Reiserer}, \citenamefont {Kalb}, \citenamefont {Blok}, \citenamefont {{van
  Bemmelen}}, \citenamefont {Taminiau}, \citenamefont {Hanson}, \citenamefont
  {Twitchen},\ and\ \citenamefont {Markham}}]{reiserer_robust_2016}%
  \BibitemOpen
  \bibfield  {author} {\bibinfo {author} {\bibfnamefont {A.}~\bibnamefont
  {Reiserer}}, \bibinfo {author} {\bibfnamefont {N.}~\bibnamefont {Kalb}},
  \bibinfo {author} {\bibfnamefont {M.~S.}\ \bibnamefont {Blok}}, \bibinfo
  {author} {\bibfnamefont {K.~J.~M.}\ \bibnamefont {{van Bemmelen}}}, \bibinfo
  {author} {\bibfnamefont {T.~H.}\ \bibnamefont {Taminiau}}, \bibinfo {author}
  {\bibfnamefont {R.}~\bibnamefont {Hanson}}, \bibinfo {author} {\bibfnamefont
  {D.~J.}\ \bibnamefont {Twitchen}},\ and\ \bibinfo {author} {\bibfnamefont
  {M.}~\bibnamefont {Markham}},\ }\bibfield  {title} {\bibinfo {title} {Robust
  {{Quantum-Network Memory Using Decoherence-Protected Subspaces}} of {{Nuclear
  Spins}}},\ }\href {https://doi.org/10.1103/PhysRevX.6.021040} {\bibfield
  {journal} {\bibinfo  {journal} {Phys. Rev. X}\ }\textbf {\bibinfo {volume}
  {6}},\ \bibinfo {pages} {021040} (\bibinfo {year} {2016})}\BibitemShut
  {NoStop}%
\bibitem [{\citenamefont {Maurer}\ \emph {et~al.}(2012)\citenamefont {Maurer},
  \citenamefont {Kucsko}, \citenamefont {Latta}, \citenamefont {Jiang},
  \citenamefont {Yao}, \citenamefont {Bennett}, \citenamefont {Pastawski},
  \citenamefont {Hunger}, \citenamefont {Chisholm}, \citenamefont {Markham},
  \citenamefont {Twitchen}, \citenamefont {Cirac},\ and\ \citenamefont
  {Lukin}}]{maurer_room-temperature_2012}%
  \BibitemOpen
  \bibfield  {author} {\bibinfo {author} {\bibfnamefont {P.~C.}\ \bibnamefont
  {Maurer}}, \bibinfo {author} {\bibfnamefont {G.}~\bibnamefont {Kucsko}},
  \bibinfo {author} {\bibfnamefont {C.}~\bibnamefont {Latta}}, \bibinfo
  {author} {\bibfnamefont {L.}~\bibnamefont {Jiang}}, \bibinfo {author}
  {\bibfnamefont {N.~Y.}\ \bibnamefont {Yao}}, \bibinfo {author} {\bibfnamefont
  {S.~D.}\ \bibnamefont {Bennett}}, \bibinfo {author} {\bibfnamefont
  {F.}~\bibnamefont {Pastawski}}, \bibinfo {author} {\bibfnamefont
  {D.}~\bibnamefont {Hunger}}, \bibinfo {author} {\bibfnamefont
  {N.}~\bibnamefont {Chisholm}}, \bibinfo {author} {\bibfnamefont
  {M.}~\bibnamefont {Markham}}, \bibinfo {author} {\bibfnamefont {D.~J.}\
  \bibnamefont {Twitchen}}, \bibinfo {author} {\bibfnamefont {J.~I.}\
  \bibnamefont {Cirac}},\ and\ \bibinfo {author} {\bibfnamefont {M.~D.}\
  \bibnamefont {Lukin}},\ }\bibfield  {title} {\bibinfo {title}
  {Room-{{Temperature Quantum Bit Memory Exceeding One Second}}},\ }\href
  {https://doi.org/10.1126/science.1220513} {\bibfield  {journal} {\bibinfo
  {journal} {Science}\ }\textbf {\bibinfo {volume} {336}},\ \bibinfo {pages}
  {1283} (\bibinfo {year} {2012})}\BibitemShut {NoStop}%
\bibitem [{\citenamefont {Doherty}\ \emph {et~al.}(2013)\citenamefont
  {Doherty}, \citenamefont {Manson}, \citenamefont {Delaney}, \citenamefont
  {Jelezko}, \citenamefont {Wrachtrup},\ and\ \citenamefont
  {Hollenberg}}]{doherty_nitrogen-vacancy_2013}%
  \BibitemOpen
  \bibfield  {author} {\bibinfo {author} {\bibfnamefont {M.~W.}\ \bibnamefont
  {Doherty}}, \bibinfo {author} {\bibfnamefont {N.~B.}\ \bibnamefont {Manson}},
  \bibinfo {author} {\bibfnamefont {P.}~\bibnamefont {Delaney}}, \bibinfo
  {author} {\bibfnamefont {F.}~\bibnamefont {Jelezko}}, \bibinfo {author}
  {\bibfnamefont {J.}~\bibnamefont {Wrachtrup}},\ and\ \bibinfo {author}
  {\bibfnamefont {L.~C.}\ \bibnamefont {Hollenberg}},\ }\bibfield  {title}
  {\bibinfo {title} {The nitrogen-vacancy colour centre in diamond},\ }\href
  {https://doi.org/10.1016/j.physrep.2013.02.001} {\bibfield  {journal}
  {\bibinfo  {journal} {Phys. Rep.}\ }\textbf {\bibinfo {volume} {528}},\
  \bibinfo {pages} {1} (\bibinfo {year} {2013})}\BibitemShut {NoStop}%
\bibitem [{\citenamefont {Li}\ \emph {et~al.}(2022)\citenamefont {Li},
  \citenamefont {Soleyman}, \citenamefont {Kohandel},\ and\ \citenamefont
  {Cappellaro}}]{Li2022}%
  \BibitemOpen
  \bibfield  {author} {\bibinfo {author} {\bibfnamefont {C.}~\bibnamefont
  {Li}}, \bibinfo {author} {\bibfnamefont {R.}~\bibnamefont {Soleyman}},
  \bibinfo {author} {\bibfnamefont {M.}~\bibnamefont {Kohandel}},\ and\
  \bibinfo {author} {\bibfnamefont {P.}~\bibnamefont {Cappellaro}},\ }\bibfield
   {title} {\bibinfo {title} {Sars-cov-2 quantum sensor based on
  nitrogen-vacancy centers in diamond},\ }\href
  {https://doi.org/10.1021/acs.nanolett.1c02868} {\bibfield  {journal}
  {\bibinfo  {journal} {Nano Lett.}\ }\textbf {\bibinfo {volume} {22}},\
  \bibinfo {pages} {43} (\bibinfo {year} {2022})}\BibitemShut {NoStop}%
\bibitem [{\citenamefont {Rendler}\ \emph {et~al.}(2017)\citenamefont
  {Rendler}, \citenamefont {Neburkova}, \citenamefont {Zemek}, \citenamefont
  {Kotek}, \citenamefont {Zappe}, \citenamefont {Chu}, \citenamefont {Cigler},\
  and\ \citenamefont {Wrachtrup}}]{Rendler_nc2017}%
  \BibitemOpen
  \bibfield  {author} {\bibinfo {author} {\bibfnamefont {T.}~\bibnamefont
  {Rendler}}, \bibinfo {author} {\bibfnamefont {J.}~\bibnamefont {Neburkova}},
  \bibinfo {author} {\bibfnamefont {O.}~\bibnamefont {Zemek}}, \bibinfo
  {author} {\bibfnamefont {J.}~\bibnamefont {Kotek}}, \bibinfo {author}
  {\bibfnamefont {A.}~\bibnamefont {Zappe}}, \bibinfo {author} {\bibfnamefont
  {Z.}~\bibnamefont {Chu}}, \bibinfo {author} {\bibfnamefont {P.}~\bibnamefont
  {Cigler}},\ and\ \bibinfo {author} {\bibfnamefont {J.}~\bibnamefont
  {Wrachtrup}},\ }\bibfield  {title} {\bibinfo {title} {Optical imaging of
  localized chemical events using programmable diamond quantum nanosensors},\
  }\href {https://doi.org/10.1038/ncomms14701} {\bibfield  {journal} {\bibinfo
  {journal} {Nat. Commun.}\ }\textbf {\bibinfo {volume} {8}},\ \bibinfo {pages}
  {14701} (\bibinfo {year} {2017})}\BibitemShut {NoStop}%
\bibitem [{\citenamefont {Pompili}\ \emph {et~al.}(2021)\citenamefont
  {Pompili}, \citenamefont {Hermans}, \citenamefont {Baier}, \citenamefont
  {Beukers}, \citenamefont {Humphreys}, \citenamefont {Schouten}, \citenamefont
  {Vermeulen}, \citenamefont {Tiggelman}, \citenamefont {dos Santos~Martins},
  \citenamefont {Dirkse}, \citenamefont {Wehner},\ and\ \citenamefont
  {Hanson}}]{PompiliScience2021}%
  \BibitemOpen
  \bibfield  {author} {\bibinfo {author} {\bibfnamefont {M.}~\bibnamefont
  {Pompili}}, \bibinfo {author} {\bibfnamefont {S.~L.~N.}\ \bibnamefont
  {Hermans}}, \bibinfo {author} {\bibfnamefont {S.}~\bibnamefont {Baier}},
  \bibinfo {author} {\bibfnamefont {H.~K.~C.}\ \bibnamefont {Beukers}},
  \bibinfo {author} {\bibfnamefont {P.~C.}\ \bibnamefont {Humphreys}}, \bibinfo
  {author} {\bibfnamefont {R.~N.}\ \bibnamefont {Schouten}}, \bibinfo {author}
  {\bibfnamefont {R.~F.~L.}\ \bibnamefont {Vermeulen}}, \bibinfo {author}
  {\bibfnamefont {M.~J.}\ \bibnamefont {Tiggelman}}, \bibinfo {author}
  {\bibfnamefont {L.}~\bibnamefont {dos Santos~Martins}}, \bibinfo {author}
  {\bibfnamefont {B.}~\bibnamefont {Dirkse}}, \bibinfo {author} {\bibfnamefont
  {S.}~\bibnamefont {Wehner}},\ and\ \bibinfo {author} {\bibfnamefont
  {R.}~\bibnamefont {Hanson}},\ }\bibfield  {title} {\bibinfo {title}
  {Realization of a multinode quantum network of remote solid-state qubits},\
  }\href {https://doi.org/10.1126/science.abg1919} {\bibfield  {journal}
  {\bibinfo  {journal} {Science}\ }\textbf {\bibinfo {volume} {372}},\ \bibinfo
  {pages} {259} (\bibinfo {year} {2021})}\BibitemShut {NoStop}%
\bibitem [{\citenamefont {Humphreys}\ \emph {et~al.}(2018)\citenamefont
  {Humphreys}, \citenamefont {Kalb}, \citenamefont {Morits}, \citenamefont
  {Schouten}, \citenamefont {Vermeulen}, \citenamefont {Twitchen},
  \citenamefont {Markham},\ and\ \citenamefont
  {Hanson}}]{Humphreys_nature2018}%
  \BibitemOpen
  \bibfield  {author} {\bibinfo {author} {\bibfnamefont {P.~C.}\ \bibnamefont
  {Humphreys}}, \bibinfo {author} {\bibfnamefont {N.}~\bibnamefont {Kalb}},
  \bibinfo {author} {\bibfnamefont {J.~P.~J.}\ \bibnamefont {Morits}}, \bibinfo
  {author} {\bibfnamefont {R.~N.}\ \bibnamefont {Schouten}}, \bibinfo {author}
  {\bibfnamefont {R.~F.~L.}\ \bibnamefont {Vermeulen}}, \bibinfo {author}
  {\bibfnamefont {D.~J.}\ \bibnamefont {Twitchen}}, \bibinfo {author}
  {\bibfnamefont {M.}~\bibnamefont {Markham}},\ and\ \bibinfo {author}
  {\bibfnamefont {R.}~\bibnamefont {Hanson}},\ }\bibfield  {title} {\bibinfo
  {title} {Deterministic delivery of remote entanglement on a quantum
  network},\ }\href {https://doi.org/10.1038/s41586-018-0200-5} {\bibfield
  {journal} {\bibinfo  {journal} {Nature}\ }\textbf {\bibinfo {volume} {558}},\
  \bibinfo {pages} {268} (\bibinfo {year} {2018})}\BibitemShut {NoStop}%
\bibitem [{\citenamefont {Fischer}\ \emph {et~al.}(2013)\citenamefont
  {Fischer}, \citenamefont {Jarmola}, \citenamefont {Kehayias},\ and\
  \citenamefont {Budker}}]{fischer_optical_2013}%
  \BibitemOpen
  \bibfield  {author} {\bibinfo {author} {\bibfnamefont {R.}~\bibnamefont
  {Fischer}}, \bibinfo {author} {\bibfnamefont {A.}~\bibnamefont {Jarmola}},
  \bibinfo {author} {\bibfnamefont {P.}~\bibnamefont {Kehayias}},\ and\
  \bibinfo {author} {\bibfnamefont {D.}~\bibnamefont {Budker}},\ }\bibfield
  {title} {\bibinfo {title} {Optical polarization of nuclear ensembles in
  diamond},\ }\href {https://doi.org/10.1103/PhysRevB.87.125207} {\bibfield
  {journal} {\bibinfo  {journal} {Phys. Rev. B}\ }\textbf {\bibinfo {volume}
  {87}},\ \bibinfo {pages} {125207} (\bibinfo {year} {2013})}\BibitemShut
  {NoStop}%
\bibitem [{\citenamefont {Chakraborty}\ \emph {et~al.}(2017)\citenamefont
  {Chakraborty}, \citenamefont {Zhang},\ and\ \citenamefont
  {Suter}}]{chakraborty_polarizing_2017}%
  \BibitemOpen
  \bibfield  {author} {\bibinfo {author} {\bibfnamefont {T.}~\bibnamefont
  {Chakraborty}}, \bibinfo {author} {\bibfnamefont {J.}~\bibnamefont {Zhang}},\
  and\ \bibinfo {author} {\bibfnamefont {D.}~\bibnamefont {Suter}},\ }\bibfield
   {title} {\bibinfo {title} {Polarizing the electronic and nuclear spin of the
  {{NV-center}} in diamond in arbitrary magnetic fields: Analysis of the
  optical pumping process},\ }\href {https://doi.org/10.1088/1367-2630/aa7727}
  {\bibfield  {journal} {\bibinfo  {journal} {New J. Phys.}\ }\textbf {\bibinfo
  {volume} {19}},\ \bibinfo {pages} {073030} (\bibinfo {year}
  {2017})}\BibitemShut {NoStop}%
\bibitem [{\citenamefont {Jaskula}\ \emph {et~al.}(2019)\citenamefont
  {Jaskula}, \citenamefont {Saha}, \citenamefont {Ajoy}, \citenamefont
  {Twitchen}, \citenamefont {Markham},\ and\ \citenamefont
  {Cappellaro}}]{jaskula_cross-sensor_2019}%
  \BibitemOpen
  \bibfield  {author} {\bibinfo {author} {\bibfnamefont {J.-C.}\ \bibnamefont
  {Jaskula}}, \bibinfo {author} {\bibfnamefont {K.}~\bibnamefont {Saha}},
  \bibinfo {author} {\bibfnamefont {A.}~\bibnamefont {Ajoy}}, \bibinfo {author}
  {\bibfnamefont {D.}~\bibnamefont {Twitchen}}, \bibinfo {author}
  {\bibfnamefont {M.}~\bibnamefont {Markham}},\ and\ \bibinfo {author}
  {\bibfnamefont {P.}~\bibnamefont {Cappellaro}},\ }\bibfield  {title}
  {\bibinfo {title} {Cross-{{Sensor Feedback Stabilization}} of an {{Emulated
  Quantum Spin Gyroscope}}},\ }\href
  {https://doi.org/10.1103/PhysRevApplied.11.054010} {\bibfield  {journal}
  {\bibinfo  {journal} {Phys. Rev. Appl.}\ }\textbf {\bibinfo {volume} {11}},\
  \bibinfo {pages} {054010} (\bibinfo {year} {2019})}\BibitemShut {NoStop}%
\bibitem [{\citenamefont {Dr{\'e}au}\ \emph {et~al.}(2013)\citenamefont
  {Dr{\'e}au}, \citenamefont {Spinicelli}, \citenamefont {Maze}, \citenamefont
  {Roch},\ and\ \citenamefont {Jacques}}]{dreau_single-shot_2013}%
  \BibitemOpen
  \bibfield  {author} {\bibinfo {author} {\bibfnamefont {A.}~\bibnamefont
  {Dr{\'e}au}}, \bibinfo {author} {\bibfnamefont {P.}~\bibnamefont
  {Spinicelli}}, \bibinfo {author} {\bibfnamefont {J.~R.}\ \bibnamefont
  {Maze}}, \bibinfo {author} {\bibfnamefont {J.-F.}\ \bibnamefont {Roch}},\
  and\ \bibinfo {author} {\bibfnamefont {V.}~\bibnamefont {Jacques}},\
  }\bibfield  {title} {\bibinfo {title} {Single-{{Shot Readout}} of {{Multiple
  Nuclear Spin Qubits}} in {{Diamond}} under {{Ambient Conditions}}},\ }\href
  {https://doi.org/10.1103/PhysRevLett.110.060502} {\bibfield  {journal}
  {\bibinfo  {journal} {Phys. Rev. Lett.}\ }\textbf {\bibinfo {volume} {110}},\
  \bibinfo {pages} {060502} (\bibinfo {year} {2013})}\BibitemShut {NoStop}%
\bibitem [{\citenamefont {Neumann}\ \emph {et~al.}(2010)\citenamefont
  {Neumann}, \citenamefont {Beck}, \citenamefont {Steiner}, \citenamefont
  {Rempp}, \citenamefont {Fedder}, \citenamefont {Hemmer}, \citenamefont
  {Wrachtrup},\ and\ \citenamefont {Jelezko}}]{neumann_single-shot_2010}%
  \BibitemOpen
  \bibfield  {author} {\bibinfo {author} {\bibfnamefont {P.}~\bibnamefont
  {Neumann}}, \bibinfo {author} {\bibfnamefont {J.}~\bibnamefont {Beck}},
  \bibinfo {author} {\bibfnamefont {M.}~\bibnamefont {Steiner}}, \bibinfo
  {author} {\bibfnamefont {F.}~\bibnamefont {Rempp}}, \bibinfo {author}
  {\bibfnamefont {H.}~\bibnamefont {Fedder}}, \bibinfo {author} {\bibfnamefont
  {P.~R.}\ \bibnamefont {Hemmer}}, \bibinfo {author} {\bibfnamefont
  {J.}~\bibnamefont {Wrachtrup}},\ and\ \bibinfo {author} {\bibfnamefont
  {F.}~\bibnamefont {Jelezko}},\ }\bibfield  {title} {\bibinfo {title}
  {Single-{{Shot Readout}} of a {{Single Nuclear Spin}}},\ }\href
  {https://doi.org/10.1126/science.1189075} {\bibfield  {journal} {\bibinfo
  {journal} {Science}\ }\textbf {\bibinfo {volume} {329}},\ \bibinfo {pages}
  {542} (\bibinfo {year} {2010})}\BibitemShut {NoStop}%
\bibitem [{\citenamefont {Liu}\ \emph {et~al.}(2017)\citenamefont {Liu},
  \citenamefont {Xing}, \citenamefont {Ma}, \citenamefont {Wang}, \citenamefont
  {Li}, \citenamefont {Po}, \citenamefont {Zhang}, \citenamefont {Fan},
  \citenamefont {Liu},\ and\ \citenamefont {Pan}}]{Liu_single_shotPRL2017}%
  \BibitemOpen
  \bibfield  {author} {\bibinfo {author} {\bibfnamefont {G.-Q.}\ \bibnamefont
  {Liu}}, \bibinfo {author} {\bibfnamefont {J.}~\bibnamefont {Xing}}, \bibinfo
  {author} {\bibfnamefont {W.-L.}\ \bibnamefont {Ma}}, \bibinfo {author}
  {\bibfnamefont {P.}~\bibnamefont {Wang}}, \bibinfo {author} {\bibfnamefont
  {C.-H.}\ \bibnamefont {Li}}, \bibinfo {author} {\bibfnamefont {H.~C.}\
  \bibnamefont {Po}}, \bibinfo {author} {\bibfnamefont {Y.-R.}\ \bibnamefont
  {Zhang}}, \bibinfo {author} {\bibfnamefont {H.}~\bibnamefont {Fan}}, \bibinfo
  {author} {\bibfnamefont {R.-B.}\ \bibnamefont {Liu}},\ and\ \bibinfo {author}
  {\bibfnamefont {X.-Y.}\ \bibnamefont {Pan}},\ }\bibfield  {title} {\bibinfo
  {title} {Single-shot readout of a nuclear spin weakly coupled to a
  nitrogen-vacancy center at room temperature},\ }\href
  {https://doi.org/10.1103/PhysRevLett.118.150504} {\bibfield  {journal}
  {\bibinfo  {journal} {Phys. Rev. Lett.}\ }\textbf {\bibinfo {volume} {118}},\
  \bibinfo {pages} {150504} (\bibinfo {year} {2017})}\BibitemShut {NoStop}%
\bibitem [{\citenamefont {Soshenko}\ \emph {et~al.}(2020)\citenamefont
  {Soshenko}, \citenamefont {Vorobyov}, \citenamefont {Bolshedvorskii},
  \citenamefont {Rubinas}, \citenamefont {Cojocaru}, \citenamefont {Kudlatsky},
  \citenamefont {Zeleneev}, \citenamefont {Sorokin}, \citenamefont
  {Smolyaninov},\ and\ \citenamefont {Akimov}}]{soshenko_temperature_2020}%
  \BibitemOpen
  \bibfield  {author} {\bibinfo {author} {\bibfnamefont {V.~V.}\ \bibnamefont
  {Soshenko}}, \bibinfo {author} {\bibfnamefont {V.~V.}\ \bibnamefont
  {Vorobyov}}, \bibinfo {author} {\bibfnamefont {S.~V.}\ \bibnamefont
  {Bolshedvorskii}}, \bibinfo {author} {\bibfnamefont {O.}~\bibnamefont
  {Rubinas}}, \bibinfo {author} {\bibfnamefont {I.}~\bibnamefont {Cojocaru}},
  \bibinfo {author} {\bibfnamefont {B.}~\bibnamefont {Kudlatsky}}, \bibinfo
  {author} {\bibfnamefont {A.~I.}\ \bibnamefont {Zeleneev}}, \bibinfo {author}
  {\bibfnamefont {V.~N.}\ \bibnamefont {Sorokin}}, \bibinfo {author}
  {\bibfnamefont {A.~N.}\ \bibnamefont {Smolyaninov}},\ and\ \bibinfo {author}
  {\bibfnamefont {A.~V.}\ \bibnamefont {Akimov}},\ }\bibfield  {title}
  {\bibinfo {title} {Temperature drift rate for nuclear terms of the
  {{NV-center}} ground-state {{Hamiltonian}}},\ }\href
  {https://doi.org/10.1103/PhysRevB.102.125133} {\bibfield  {journal} {\bibinfo
   {journal} {Phys. Rev. B}\ }\textbf {\bibinfo {volume} {102}},\ \bibinfo
  {pages} {125133} (\bibinfo {year} {2020})}\BibitemShut {NoStop}%
\bibitem [{\citenamefont {Jarmola}\ \emph {et~al.}(2020)\citenamefont
  {Jarmola}, \citenamefont {Fescenko}, \citenamefont {Acosta}, \citenamefont
  {Doherty}, \citenamefont {Fatemi}, \citenamefont {Ivanov}, \citenamefont
  {Budker},\ and\ \citenamefont {Malinovsky}}]{jarmola_robust_2020}%
  \BibitemOpen
  \bibfield  {author} {\bibinfo {author} {\bibfnamefont {A.}~\bibnamefont
  {Jarmola}}, \bibinfo {author} {\bibfnamefont {I.}~\bibnamefont {Fescenko}},
  \bibinfo {author} {\bibfnamefont {V.~M.}\ \bibnamefont {Acosta}}, \bibinfo
  {author} {\bibfnamefont {M.~W.}\ \bibnamefont {Doherty}}, \bibinfo {author}
  {\bibfnamefont {F.~K.}\ \bibnamefont {Fatemi}}, \bibinfo {author}
  {\bibfnamefont {T.}~\bibnamefont {Ivanov}}, \bibinfo {author} {\bibfnamefont
  {D.}~\bibnamefont {Budker}},\ and\ \bibinfo {author} {\bibfnamefont {V.~S.}\
  \bibnamefont {Malinovsky}},\ }\bibfield  {title} {\bibinfo {title} {Robust
  optical readout and characterization of nuclear spin transitions in
  nitrogen-vacancy ensembles in diamond},\ }\href
  {https://doi.org/10.1103/PhysRevResearch.2.023094} {\bibfield  {journal}
  {\bibinfo  {journal} {Phys. Rev. Res.}\ }\textbf {\bibinfo {volume} {2}},\
  \bibinfo {pages} {023094} (\bibinfo {year} {2020})}\BibitemShut {NoStop}%
\bibitem [{Note1()}]{Note1}%
  \BibitemOpen
  \bibinfo {note} {In general, the decay is related to the p.d.f.
  characteristic functions of the temperature and strain
  variations.}\BibitemShut {Stop}%
\bibitem [{\citenamefont {Doherty}\ \emph
  {et~al.}(2014{\natexlab{b}})\citenamefont {Doherty}, \citenamefont {Acosta},
  \citenamefont {Jarmola}, \citenamefont {Barson}, \citenamefont {Manson},
  \citenamefont {Budker},\ and\ \citenamefont
  {Hollenberg}}]{doherty_temperature_2014}%
  \BibitemOpen
  \bibfield  {author} {\bibinfo {author} {\bibfnamefont {M.~W.}\ \bibnamefont
  {Doherty}}, \bibinfo {author} {\bibfnamefont {V.~M.}\ \bibnamefont {Acosta}},
  \bibinfo {author} {\bibfnamefont {A.}~\bibnamefont {Jarmola}}, \bibinfo
  {author} {\bibfnamefont {M.~S.~J.}\ \bibnamefont {Barson}}, \bibinfo {author}
  {\bibfnamefont {N.~B.}\ \bibnamefont {Manson}}, \bibinfo {author}
  {\bibfnamefont {D.}~\bibnamefont {Budker}},\ and\ \bibinfo {author}
  {\bibfnamefont {L.~C.~L.}\ \bibnamefont {Hollenberg}},\ }\bibfield  {title}
  {\bibinfo {title} {Temperature shifts of the resonances of the {{NV}} -
  center in diamond},\ }\href {https://doi.org/10.1103/PhysRevB.90.041201}
  {\bibfield  {journal} {\bibinfo  {journal} {Phys. Rev. B}\ }\textbf {\bibinfo
  {volume} {90}},\ \bibinfo {pages} {041201} (\bibinfo {year}
  {2014}{\natexlab{b}})}\BibitemShut {NoStop}%
\bibitem [{\citenamefont {Tang}\ \emph {et~al.}(2022)\citenamefont {Tang},
  \citenamefont {Barr}, \citenamefont {Wang}, \citenamefont {Cappellaro},\ and\
  \citenamefont {Li}}]{Tang_Temperature_2022}%
  \BibitemOpen
  \bibfield  {author} {\bibinfo {author} {\bibfnamefont {H.}~\bibnamefont
  {Tang}}, \bibinfo {author} {\bibfnamefont {A.~R.}\ \bibnamefont {Barr}},
  \bibinfo {author} {\bibfnamefont {G.}~\bibnamefont {Wang}}, \bibinfo {author}
  {\bibfnamefont {P.}~\bibnamefont {Cappellaro}},\ and\ \bibinfo {author}
  {\bibfnamefont {J.}~\bibnamefont {Li}},\ }\href@noop {} {\bibinfo {title}
  {First-principles calculation of the temperature-dependent transition
  energies in spin defects}} (\bibinfo {year} {2022}),\ \Eprint
  {https://arxiv.org/abs/arXiv:2205.02791} {arXiv:2205.02791} \BibitemShut
  {NoStop}%
\bibitem [{\citenamefont {Shang}\ \emph {et~al.}(2022)\citenamefont {Shang},
  \citenamefont {Hong}, \citenamefont {Dai}, \citenamefont {Lu}, \citenamefont
  {Yu}, \citenamefont {Yu}, \citenamefont {Yu}, \citenamefont {Pan},\ and\
  \citenamefont {Liu}}]{arxiv2203.10511}%
  \BibitemOpen
  \bibfield  {author} {\bibinfo {author} {\bibfnamefont {Y.-X.}\ \bibnamefont
  {Shang}}, \bibinfo {author} {\bibfnamefont {F.}~\bibnamefont {Hong}},
  \bibinfo {author} {\bibfnamefont {J.-H.}\ \bibnamefont {Dai}}, \bibinfo
  {author} {\bibfnamefont {Y.-N.}\ \bibnamefont {Lu}}, \bibinfo {author}
  {\bibfnamefont {H.}~\bibnamefont {Yu}}, \bibinfo {author} {\bibfnamefont
  {Y.-H.}\ \bibnamefont {Yu}}, \bibinfo {author} {\bibfnamefont {X.-H.}\
  \bibnamefont {Yu}}, \bibinfo {author} {\bibfnamefont {X.-Y.}\ \bibnamefont
  {Pan}},\ and\ \bibinfo {author} {\bibfnamefont {G.-Q.}\ \bibnamefont {Liu}},\
  }\href@noop {} {\bibinfo {title} {High-pressure nmr enabled by diamond
  nitrogen-vacancy centers}} (\bibinfo {year} {2022}),\ \Eprint
  {https://arxiv.org/abs/2203.10511} {arXiv:2203.10511 [quant-ph]} \BibitemShut
  {NoStop}%
\bibitem [{\citenamefont {Wang}\ \emph {et~al.}(2021)\citenamefont {Wang},
  \citenamefont {McPherson}, \citenamefont {Kadado}, \citenamefont {Brandt},
  \citenamefont {Edwards}, \citenamefont {Casey},\ and\ \citenamefont
  {Curro}}]{wang_ac_2021}%
  \BibitemOpen
  \bibfield  {author} {\bibinfo {author} {\bibfnamefont {Z.}~\bibnamefont
  {Wang}}, \bibinfo {author} {\bibfnamefont {C.}~\bibnamefont {McPherson}},
  \bibinfo {author} {\bibfnamefont {R.}~\bibnamefont {Kadado}}, \bibinfo
  {author} {\bibfnamefont {N.}~\bibnamefont {Brandt}}, \bibinfo {author}
  {\bibfnamefont {S.}~\bibnamefont {Edwards}}, \bibinfo {author} {\bibfnamefont
  {W.}~\bibnamefont {Casey}},\ and\ \bibinfo {author} {\bibfnamefont
  {N.}~\bibnamefont {Curro}},\ }\bibfield  {title} {\bibinfo {title} {Ac
  {{Sensing Using Nitrogen-Vacancy Centers}} in a {{Diamond Anvil Cell}} up to
  6 {{GPa}}},\ }\href {https://doi.org/10.1103/PhysRevApplied.16.054014}
  {\bibfield  {journal} {\bibinfo  {journal} {Phys. Rev. Appl.}\ }\textbf
  {\bibinfo {volume} {16}},\ \bibinfo {pages} {054014} (\bibinfo {year}
  {2021})}\BibitemShut {NoStop}%
\bibitem [{\citenamefont {Udvarhelyi}\ \emph {et~al.}(2018)\citenamefont
  {Udvarhelyi}, \citenamefont {Shkolnikov}, \citenamefont {Gali}, \citenamefont
  {Burkard},\ and\ \citenamefont {P{\'a}lyi}}]{udvarhelyi_spin-strain_2018-1}%
  \BibitemOpen
  \bibfield  {author} {\bibinfo {author} {\bibfnamefont {P.}~\bibnamefont
  {Udvarhelyi}}, \bibinfo {author} {\bibfnamefont {V.~O.}\ \bibnamefont
  {Shkolnikov}}, \bibinfo {author} {\bibfnamefont {A.}~\bibnamefont {Gali}},
  \bibinfo {author} {\bibfnamefont {G.}~\bibnamefont {Burkard}},\ and\ \bibinfo
  {author} {\bibfnamefont {A.}~\bibnamefont {P{\'a}lyi}},\ }\bibfield  {title}
  {\bibinfo {title} {Spin-strain interaction in nitrogen-vacancy centers in
  diamond},\ }\href {https://doi.org/10.1103/PhysRevB.98.075201} {\bibfield
  {journal} {\bibinfo  {journal} {Phys. Rev. B}\ }\textbf {\bibinfo {volume}
  {98}},\ \bibinfo {pages} {075201} (\bibinfo {year} {2018})}\BibitemShut
  {NoStop}%
\bibitem [{\citenamefont {Meesala}\ \emph {et~al.}(2018)\citenamefont
  {Meesala}, \citenamefont {Sohn}, \citenamefont {Pingault}, \citenamefont
  {Shao}, \citenamefont {Atikian}, \citenamefont {Holzgrafe}, \citenamefont
  {G{\"u}ndo{\u g}an}, \citenamefont {Stavrakas}, \citenamefont {Sipahigil},
  \citenamefont {Chia}, \citenamefont {Evans}, \citenamefont {Burek},
  \citenamefont {Zhang}, \citenamefont {Wu}, \citenamefont {Pacheco},
  \citenamefont {Abraham}, \citenamefont {Bielejec}, \citenamefont {Lukin},
  \citenamefont {Atat{\"u}re},\ and\ \citenamefont {Lon{\v
  c}ar}}]{meesala_strain_2018}%
  \BibitemOpen
  \bibfield  {author} {\bibinfo {author} {\bibfnamefont {S.}~\bibnamefont
  {Meesala}}, \bibinfo {author} {\bibfnamefont {Y.-I.}\ \bibnamefont {Sohn}},
  \bibinfo {author} {\bibfnamefont {B.}~\bibnamefont {Pingault}}, \bibinfo
  {author} {\bibfnamefont {L.}~\bibnamefont {Shao}}, \bibinfo {author}
  {\bibfnamefont {H.~A.}\ \bibnamefont {Atikian}}, \bibinfo {author}
  {\bibfnamefont {J.}~\bibnamefont {Holzgrafe}}, \bibinfo {author}
  {\bibfnamefont {M.}~\bibnamefont {G{\"u}ndo{\u g}an}}, \bibinfo {author}
  {\bibfnamefont {C.}~\bibnamefont {Stavrakas}}, \bibinfo {author}
  {\bibfnamefont {A.}~\bibnamefont {Sipahigil}}, \bibinfo {author}
  {\bibfnamefont {C.}~\bibnamefont {Chia}}, \bibinfo {author} {\bibfnamefont
  {R.}~\bibnamefont {Evans}}, \bibinfo {author} {\bibfnamefont {M.~J.}\
  \bibnamefont {Burek}}, \bibinfo {author} {\bibfnamefont {M.}~\bibnamefont
  {Zhang}}, \bibinfo {author} {\bibfnamefont {L.}~\bibnamefont {Wu}}, \bibinfo
  {author} {\bibfnamefont {J.~L.}\ \bibnamefont {Pacheco}}, \bibinfo {author}
  {\bibfnamefont {J.}~\bibnamefont {Abraham}}, \bibinfo {author} {\bibfnamefont
  {E.}~\bibnamefont {Bielejec}}, \bibinfo {author} {\bibfnamefont {M.~D.}\
  \bibnamefont {Lukin}}, \bibinfo {author} {\bibfnamefont {M.}~\bibnamefont
  {Atat{\"u}re}},\ and\ \bibinfo {author} {\bibfnamefont {M.}~\bibnamefont
  {Lon{\v c}ar}},\ }\bibfield  {title} {\bibinfo {title} {Strain engineering of
  the silicon-vacancy center in diamond},\ }\href
  {https://doi.org/10.1103/PhysRevB.97.205444} {\bibfield  {journal} {\bibinfo
  {journal} {Phys. Rev. B}\ }\textbf {\bibinfo {volume} {97}},\ \bibinfo
  {pages} {205444} (\bibinfo {year} {2018})}\BibitemShut {NoStop}%
\bibitem [{\citenamefont {Mansir}\ \emph {et~al.}(2018)\citenamefont {Mansir},
  \citenamefont {Conti}, \citenamefont {Zeng}, \citenamefont {Pla},
  \citenamefont {Bertet}, \citenamefont {Swift}, \citenamefont {{Van de
  Walle}}, \citenamefont {Thewalt}, \citenamefont {Sklenard}, \citenamefont
  {Niquet},\ and\ \citenamefont {Morton}}]{mansir_linear_2018}%
  \BibitemOpen
  \bibfield  {author} {\bibinfo {author} {\bibfnamefont {J.}~\bibnamefont
  {Mansir}}, \bibinfo {author} {\bibfnamefont {P.}~\bibnamefont {Conti}},
  \bibinfo {author} {\bibfnamefont {Z.}~\bibnamefont {Zeng}}, \bibinfo {author}
  {\bibfnamefont {J.~J.}\ \bibnamefont {Pla}}, \bibinfo {author} {\bibfnamefont
  {P.}~\bibnamefont {Bertet}}, \bibinfo {author} {\bibfnamefont {M.~W.}\
  \bibnamefont {Swift}}, \bibinfo {author} {\bibfnamefont {C.~G.}\ \bibnamefont
  {{Van de Walle}}}, \bibinfo {author} {\bibfnamefont {M.~L.~W.}\ \bibnamefont
  {Thewalt}}, \bibinfo {author} {\bibfnamefont {B.}~\bibnamefont {Sklenard}},
  \bibinfo {author} {\bibfnamefont {Y.~M.}\ \bibnamefont {Niquet}},\ and\
  \bibinfo {author} {\bibfnamefont {J.~J.~L.}\ \bibnamefont {Morton}},\
  }\bibfield  {title} {\bibinfo {title} {Linear {{Hyperfine Tuning}} of {{Donor
  Spins}} in {{Silicon Using Hydrostatic Strain}}},\ }\href
  {https://doi.org/10.1103/PhysRevLett.120.167701} {\bibfield  {journal}
  {\bibinfo  {journal} {Phys. Rev. Lett.}\ }\textbf {\bibinfo {volume} {120}},\
  \bibinfo {pages} {167701} (\bibinfo {year} {2018})}\BibitemShut {NoStop}%
\bibitem [{\citenamefont {Wolfowicz}\ \emph {et~al.}(2014)\citenamefont
  {Wolfowicz}, \citenamefont {Urdampilleta}, \citenamefont {Thewalt},
  \citenamefont {Riemann}, \citenamefont {Abrosimov}, \citenamefont {Becker},
  \citenamefont {Pohl},\ and\ \citenamefont
  {Morton}}]{wolfowicz_conditional_2014}%
  \BibitemOpen
  \bibfield  {author} {\bibinfo {author} {\bibfnamefont {G.}~\bibnamefont
  {Wolfowicz}}, \bibinfo {author} {\bibfnamefont {M.}~\bibnamefont
  {Urdampilleta}}, \bibinfo {author} {\bibfnamefont {M.~L.~W.}\ \bibnamefont
  {Thewalt}}, \bibinfo {author} {\bibfnamefont {H.}~\bibnamefont {Riemann}},
  \bibinfo {author} {\bibfnamefont {N.~V.}\ \bibnamefont {Abrosimov}}, \bibinfo
  {author} {\bibfnamefont {P.}~\bibnamefont {Becker}}, \bibinfo {author}
  {\bibfnamefont {H.-J.}\ \bibnamefont {Pohl}},\ and\ \bibinfo {author}
  {\bibfnamefont {J.~J.~L.}\ \bibnamefont {Morton}},\ }\bibfield  {title}
  {\bibinfo {title} {Conditional {{Control}} of {{Donor Nuclear Spins}} in
  {{Silicon Using Stark Shifts}}},\ }\href
  {https://doi.org/10.1103/PhysRevLett.113.157601} {\bibfield  {journal}
  {\bibinfo  {journal} {Phys. Rev. Lett.}\ }\textbf {\bibinfo {volume} {113}},\
  \bibinfo {pages} {157601} (\bibinfo {year} {2014})}\BibitemShut {NoStop}%
\bibitem [{\citenamefont {Franke}\ \emph {et~al.}(2015)\citenamefont {Franke},
  \citenamefont {Hrubesch}, \citenamefont {K{\"u}nzl}, \citenamefont {Becker},
  \citenamefont {Itoh}, \citenamefont {Stutzmann}, \citenamefont {Hoehne},
  \citenamefont {Dreher},\ and\ \citenamefont
  {Brandt}}]{franke_interaction_2015}%
  \BibitemOpen
  \bibfield  {author} {\bibinfo {author} {\bibfnamefont {D.~P.}\ \bibnamefont
  {Franke}}, \bibinfo {author} {\bibfnamefont {F.~M.}\ \bibnamefont
  {Hrubesch}}, \bibinfo {author} {\bibfnamefont {M.}~\bibnamefont {K{\"u}nzl}},
  \bibinfo {author} {\bibfnamefont {H.-W.}\ \bibnamefont {Becker}}, \bibinfo
  {author} {\bibfnamefont {K.~M.}\ \bibnamefont {Itoh}}, \bibinfo {author}
  {\bibfnamefont {M.}~\bibnamefont {Stutzmann}}, \bibinfo {author}
  {\bibfnamefont {F.}~\bibnamefont {Hoehne}}, \bibinfo {author} {\bibfnamefont
  {L.}~\bibnamefont {Dreher}},\ and\ \bibinfo {author} {\bibfnamefont {M.~S.}\
  \bibnamefont {Brandt}},\ }\bibfield  {title} {\bibinfo {title} {Interaction
  of {{Strain}} and {{Nuclear Spins}} in {{Silicon}}: {{Quadrupolar Effects}}
  on {{Ionized Donors}}},\ }\href
  {https://doi.org/10.1103/PhysRevLett.115.057601} {\bibfield  {journal}
  {\bibinfo  {journal} {Phys. Rev. Lett.}\ }\textbf {\bibinfo {volume} {115}},\
  \bibinfo {pages} {057601} (\bibinfo {year} {2015})}\BibitemShut {NoStop}%
\bibitem [{SOM()}]{SOM}%
  \BibitemOpen
  \href@noop {} {}\bibinfo {howpublished} {See Supplemental Material for
  details.}\BibitemShut {Stop}%
\bibitem [{\citenamefont {Acosta}\ \emph {et~al.}(2010)\citenamefont {Acosta},
  \citenamefont {Bauch}, \citenamefont {Ledbetter}, \citenamefont {Waxman},
  \citenamefont {Bouchard},\ and\ \citenamefont
  {Budker}}]{acosta_temperature_2010}%
  \BibitemOpen
  \bibfield  {author} {\bibinfo {author} {\bibfnamefont {V.~M.}\ \bibnamefont
  {Acosta}}, \bibinfo {author} {\bibfnamefont {E.}~\bibnamefont {Bauch}},
  \bibinfo {author} {\bibfnamefont {M.~P.}\ \bibnamefont {Ledbetter}}, \bibinfo
  {author} {\bibfnamefont {A.}~\bibnamefont {Waxman}}, \bibinfo {author}
  {\bibfnamefont {L.-S.}\ \bibnamefont {Bouchard}},\ and\ \bibinfo {author}
  {\bibfnamefont {D.}~\bibnamefont {Budker}},\ }\bibfield  {title} {\bibinfo
  {title} {Temperature {{Dependence}} of the {{Nitrogen-Vacancy Magnetic
  Resonance}} in {{Diamond}}},\ }\href
  {https://doi.org/10.1103/PhysRevLett.104.070801} {\bibfield  {journal}
  {\bibinfo  {journal} {Phys. Rev. Lett.}\ }\textbf {\bibinfo {volume} {104}},\
  \bibinfo {pages} {070801} (\bibinfo {year} {2010})}\BibitemShut {NoStop}%
\bibitem [{\citenamefont {Barson}\ \emph {et~al.}(2019)\citenamefont {Barson},
  \citenamefont {Reddy}, \citenamefont {Yang}, \citenamefont {Manson},
  \citenamefont {Wrachtrup},\ and\ \citenamefont
  {Doherty}}]{barson_temperature_2019}%
  \BibitemOpen
  \bibfield  {author} {\bibinfo {author} {\bibfnamefont {M.~S.~J.}\
  \bibnamefont {Barson}}, \bibinfo {author} {\bibfnamefont {P.}~\bibnamefont
  {Reddy}}, \bibinfo {author} {\bibfnamefont {S.}~\bibnamefont {Yang}},
  \bibinfo {author} {\bibfnamefont {N.~B.}\ \bibnamefont {Manson}}, \bibinfo
  {author} {\bibfnamefont {J.}~\bibnamefont {Wrachtrup}},\ and\ \bibinfo
  {author} {\bibfnamefont {M.~W.}\ \bibnamefont {Doherty}},\ }\bibfield
  {title} {\bibinfo {title} {Temperature dependence of the {{C}} 13 hyperfine
  structure of the negatively charged nitrogen-vacancy center in diamond},\
  }\href {https://doi.org/10.1103/PhysRevB.99.094101} {\bibfield  {journal}
  {\bibinfo  {journal} {Phys. Rev. B}\ }\textbf {\bibinfo {volume} {99}},\
  \bibinfo {pages} {094101} (\bibinfo {year} {2019})}\BibitemShut {NoStop}%
\bibitem [{\citenamefont {Iv{\'a}dy}\ \emph {et~al.}(2014)\citenamefont
  {Iv{\'a}dy}, \citenamefont {Simon}, \citenamefont {Maze}, \citenamefont
  {Abrikosov},\ and\ \citenamefont {Gali}}]{ivady_pressure_2014}%
  \BibitemOpen
  \bibfield  {author} {\bibinfo {author} {\bibfnamefont {V.}~\bibnamefont
  {Iv{\'a}dy}}, \bibinfo {author} {\bibfnamefont {T.}~\bibnamefont {Simon}},
  \bibinfo {author} {\bibfnamefont {J.~R.}\ \bibnamefont {Maze}}, \bibinfo
  {author} {\bibfnamefont {I.~A.}\ \bibnamefont {Abrikosov}},\ and\ \bibinfo
  {author} {\bibfnamefont {A.}~\bibnamefont {Gali}},\ }\bibfield  {title}
  {\bibinfo {title} {Pressure and temperature dependence of the zero-field
  splitting in the ground state of {{NV}} centers in diamond: {{A}}
  first-principles study},\ }\href {https://doi.org/10.1103/PhysRevB.90.235205}
  {\bibfield  {journal} {\bibinfo  {journal} {Phys. Rev. B}\ }\textbf {\bibinfo
  {volume} {90}},\ \bibinfo {pages} {235205} (\bibinfo {year}
  {2014})}\BibitemShut {NoStop}%
\bibitem [{\citenamefont {Kresse}\ and\ \citenamefont
  {Furthm{\"u}ller}(1996)}]{kresse1996efficient}%
  \BibitemOpen
  \bibfield  {author} {\bibinfo {author} {\bibfnamefont {G.}~\bibnamefont
  {Kresse}}\ and\ \bibinfo {author} {\bibfnamefont {J.}~\bibnamefont
  {Furthm{\"u}ller}},\ }\bibfield  {title} {\bibinfo {title} {Efficient
  iterative schemes for ab initio total-energy calculations using a plane-wave
  basis set},\ }\href {https://doi.org/10.1103/PhysRevB.54.11169} {\bibfield
  {journal} {\bibinfo  {journal} {Phys. Rev. B}\ }\textbf {\bibinfo {volume}
  {54}},\ \bibinfo {pages} {11169} (\bibinfo {year} {1996})}\BibitemShut
  {NoStop}%
\bibitem [{\citenamefont {Kresse}\ and\ \citenamefont
  {Joubert}(1999)}]{kresse1999ultrasoft}%
  \BibitemOpen
  \bibfield  {author} {\bibinfo {author} {\bibfnamefont {G.}~\bibnamefont
  {Kresse}}\ and\ \bibinfo {author} {\bibfnamefont {D.}~\bibnamefont
  {Joubert}},\ }\bibfield  {title} {\bibinfo {title} {From ultrasoft
  pseudopotentials to the projector augmented-wave method},\ }\href
  {https://doi.org/10.1103/PhysRevB.59.1758} {\bibfield  {journal} {\bibinfo
  {journal} {Phys. Rev. B}\ }\textbf {\bibinfo {volume} {59}},\ \bibinfo
  {pages} {1758} (\bibinfo {year} {1999})}\BibitemShut {NoStop}%
\bibitem [{\citenamefont {Perdew}\ \emph {et~al.}(1996)\citenamefont {Perdew},
  \citenamefont {Burke},\ and\ \citenamefont
  {Ernzerhof}}]{perdew1996generalized}%
  \BibitemOpen
  \bibfield  {author} {\bibinfo {author} {\bibfnamefont {J.~P.}\ \bibnamefont
  {Perdew}}, \bibinfo {author} {\bibfnamefont {K.}~\bibnamefont {Burke}},\ and\
  \bibinfo {author} {\bibfnamefont {M.}~\bibnamefont {Ernzerhof}},\ }\bibfield
  {title} {\bibinfo {title} {Generalized gradient approximation made simple},\
  }\href {https://doi.org/10.1103/PhysRevLett.77.3865} {\bibfield  {journal}
  {\bibinfo  {journal} {Phys. Rev. Lett.}\ }\textbf {\bibinfo {volume} {77}},\
  \bibinfo {pages} {3865} (\bibinfo {year} {1996})}\BibitemShut {NoStop}%
\bibitem [{\citenamefont {Bauch}\ \emph {et~al.}(2020)\citenamefont {Bauch},
  \citenamefont {Singh}, \citenamefont {Lee}, \citenamefont {Hart},
  \citenamefont {Schloss}, \citenamefont {Turner}, \citenamefont {Barry},
  \citenamefont {Pham}, \citenamefont {{Bar-Gill}}, \citenamefont {Yelin},\
  and\ \citenamefont {Walsworth}}]{bauch_decoherence_2020}%
  \BibitemOpen
  \bibfield  {author} {\bibinfo {author} {\bibfnamefont {E.}~\bibnamefont
  {Bauch}}, \bibinfo {author} {\bibfnamefont {S.}~\bibnamefont {Singh}},
  \bibinfo {author} {\bibfnamefont {J.}~\bibnamefont {Lee}}, \bibinfo {author}
  {\bibfnamefont {C.~A.}\ \bibnamefont {Hart}}, \bibinfo {author}
  {\bibfnamefont {J.~M.}\ \bibnamefont {Schloss}}, \bibinfo {author}
  {\bibfnamefont {M.~J.}\ \bibnamefont {Turner}}, \bibinfo {author}
  {\bibfnamefont {J.~F.}\ \bibnamefont {Barry}}, \bibinfo {author}
  {\bibfnamefont {L.~M.}\ \bibnamefont {Pham}}, \bibinfo {author}
  {\bibfnamefont {N.}~\bibnamefont {{Bar-Gill}}}, \bibinfo {author}
  {\bibfnamefont {S.~F.}\ \bibnamefont {Yelin}},\ and\ \bibinfo {author}
  {\bibfnamefont {R.~L.}\ \bibnamefont {Walsworth}},\ }\bibfield  {title}
  {\bibinfo {title} {Decoherence of ensembles of nitrogen-vacancy centers in
  diamond},\ }\href {https://doi.org/10.1103/PhysRevB.102.134210} {\bibfield
  {journal} {\bibinfo  {journal} {Phys. Rev. B}\ }\textbf {\bibinfo {volume}
  {102}},\ \bibinfo {pages} {134210} (\bibinfo {year} {2020})}\BibitemShut
  {NoStop}%
\end{thebibliography}%

\end{document}